\documentclass[aps,prx,reprint,superscriptaddress,longbibliography,floatfix,nofootinbib]{revtex4-2}
\usepackage{amsmath,amssymb,amsthm,bm,mathtools}
\usepackage{bbm}
\usepackage{braket}
\usepackage{graphicx}
\usepackage{xcolor}
\usepackage[colorlinks=true,linkcolor=blue!50!black,citecolor=blue!50!black,urlcolor=blue!50!black]{hyperref}

\theoremstyle{plain}
\newtheorem{theorem}{Theorem}

\newtheorem{lemma}{Lemma}

\theoremstyle{definition}
\newtheorem{definition}{Definition}
\newtheorem{assumption}{Assumption}
\theoremstyle{remark}

\newcommand{\one}{\mathbbm{1}}
\newcommand{\dd}{\mathrm{d}}
\newcommand{\Tr}{\operatorname{Tr}}
\newcommand{\E}{\mathbb{E}}
\newcommand{\Heff}{H_{\mathrm{eff}}}

\newcommand{\Mem}{\mathcal{M}}
\newcommand{\Out}{\mathcal{I}}
\newcommand{\Pst}{\mathcal{P}}
\newcommand{\Reach}{\mathcal{R}}
\newcommand{\dunion}{\bigsqcup}

\begin{document}

\title{Optimal Controlled Quantum Instrument for Monitored Many-Body Dynamics}

\author{Jimin Li}
\email{jl939physics@gmail.com}
\affiliation{Centre for Quantum Information and Foundations, Department of Applied Mathematics and Theoretical Physics,
University of Cambridge, Wilberforce Road, Cambridge CB3 0WA, United Kingdom}

\begin{abstract}
Post-selecting rare trajectories of monitored quantum dynamics requires exponentially many repeated experiments. We extend Bayes' rule to monitored many-body dynamics and bias the quantum trajectories by nonlinear functions of the post-measurement state. For functions that are constant on the sets of a suitable finite partition of the post-measurement states, we construct a quantum instrument on the system and an additional register that produces the biased trajectories at the optimal rates without post-selection. This optimal controlled quantum instrument has a natural tensor-network representation as the left canonical form of the system--register tensors. We illustrate the construction with two examples: a precessing qubit, biased by its spin component along the field, and free fermions with gain and loss, biased by the particle number, where the biased trajectories undergo a phase transition. We also argue, under one assumption, that biasing a measured Majorana chain by its von Neumann entropy leads to entanglement transitions.
\end{abstract}

\maketitle

\section{Introduction}\label{sec:intro}

\begin{figure*}[t]
\centering
\includegraphics{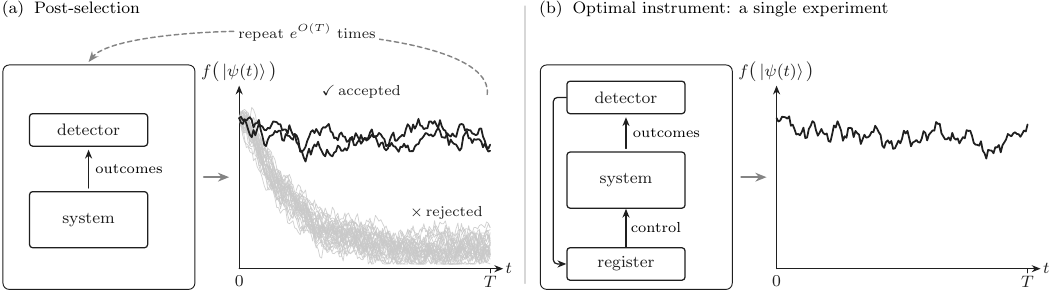}
\caption{Post-selection versus the optimal controlled quantum instrument. (a)~Post-selection accepts the rare trajectories along which a function $f$ of the post-measurement state stays large (black) and rejects the typical ones (grey), so that the experiment must be repeated a number of times that grows exponentially in $T$. (b)~The optimal instrument adds a register, which is updated by the measurement outcomes and controls the system, and produces such a trajectory in a single experiment.}
\label{fig:overview}
\end{figure*}

Monitoring and controlling quantum dynamics are fundamental parts of quantum theory, as well as of experimental platforms such as quantum circuits with mid-circuit measurements and open systems whose environment is continuously observed~\cite{WM09,Car93,DCM92,PK98}. We use quantum theory to operate quantum systems, such that the outcomes of measurements determine the operations that are applied next~\cite{WM09}. Formally, a measurement together with its effect on the state is described by a quantum instrument. Quantum theory requires a quantum instrument to be linear, and every quantum instrument can be implemented through a unitary dilation~\cite{WM09}.

Averaged over measurement outcomes, the state of a continuously monitored system evolves under the completely positive (CP), trace-preserving (TP) maps $e^{t\mathcal L}$ generated by the Lindbladian
\begin{equation}\label{eq:intro-lindblad}
  \mathcal L(\rho)\coloneqq-i[H,\rho]+\sum_\mu\Big(L_\mu\rho L_\mu^\dagger-\tfrac12\big\{L_\mu^\dagger L_\mu,\rho\big\}\Big),
\end{equation}
with Hamiltonian $H$ and jump operators $L_\mu$. Resolving the measurement outcomes gives the quantum-jump unravelling of Eq.~\eqref{eq:intro-lindblad}, in which the system undergoes a quantum instrument in every time step $\dd t$. Up to a time $T$, the system and the record of the outcomes are in the joint state
\begin{multline}\label{eq:intro-traj}
  \mathcal T\exp\Big(\int_0^T\!\dd t\,\Big(-i\Heff\otimes\one_{\mathrm{rec}}+\sum_\mu L_\mu\otimes a_\mu^\dagger(t)\Big)\Big)\\
  \times\ket{\psi_0}\ket\varnothing,
\end{multline}
where $\Heff\coloneqq H-\frac i2\sum_\mu L_\mu^\dagger L_\mu$ is the non-Hermitian effective Hamiltonian, $\ket\varnothing$ is the record with outcome $\varnothing$ at all times, $a_\mu^\dagger(t)$ adds the outcome $\mu$ at time $t$ to the record, with $[a_\mu(t),a_\nu^\dagger(t')]=\delta_{\mu\nu}\delta(t-t')$,
$\one_{\mathrm{rec}}$ is the identity of the record, and $\mathcal T$ orders the times from right to left. This is a continuous matrix product state (cMPS)~\cite{VC10,OEV10}, whose sites are the time
steps, whose physical index is the measurement outcome of each step, and whose auxiliary space is the system. Its component along a record is the
unnormalised post-measurement state, whose squared norm is the Born probability density of the record, and tracing out the record gives
$e^{T\mathcal L}(\ket{\psi_0}\bra{\psi_0})$. Individual records, however, contain much more information than this average. In particular, the mean of a nonlinear function of the post-measurement state is not a function of the averaged state.

Accessing such nonlinear properties along specific trajectories, or sets of trajectories, amounts to post-selection, and the trajectories of interest are often atypical. One may wish, for instance, to follow trajectories along which the post-measurement state remains entangled although the
measurements tend to disentangle it, or remains mixed although they tend to purify it. Such rare trajectories occur with a probability that decays
exponentially with the duration of the experiment and, in many-body systems, also with the system size, which makes direct post-selection impractical beyond small systems and short times, see Fig.~\ref{fig:overview}(a).

Measurement-induced entanglement transitions in monitored many-body dynamics were found in hybrid
circuits of unitary gates and measurements~\cite{LCF18,SRN19,CNPS19,LCF19} and have been analysed through mappings to statistical mechanics,
quantum error correction and purification~\cite{CBQA20,BCA20,JYVL20,ZGWGHP20,GH20,GH20b,SKGG20,VSGDDGH22}; see Refs.~\cite{FKNV23,PV22} for reviews. Related
transitions occur in measurement-only dynamics~\cite{LB20,NS20,IGGHK21,LAB21,SH21} and in monitored free fermions~\cite{CTD19,ABD21}, and they
have been observed on trapped-ion and superconducting processors~\cite{Noe22,KSMM23,GQAI23}. In all these settings post-selection is a central
obstacle, which has motivated probes that combine the measurement record with classical computation~\cite{GH20b,BAPGV22,LZGAF23,LJBG25,CDCBLC26}, as well as
adaptive circuits in which measurement outcomes are fed back to steer the dynamics~\cite{IGPW23,ST23,OMGK24}. In parallel, the counting
statistics of the measurement outcomes of open quantum systems have been studied~\cite{GL10,LHGG13,CGLP18,CJG19,CBL21,LBGJ26,CCCLB26}, recently also on quantum
hardware~\cite{CLC23}, as has the choice of the unravelling~\cite{BMJ25}.

Here we connect these lines of research by asking whether the unphysical statistics of a post-selected process can be
produced by a physical quantum instrument, i.e., by one whose typical trajectories are the post-selected ones. Extending Bayes' rule to quantum dynamics is essential here: given that a trajectory will eventually be accepted, the probability of the measurement outcome $\mu$ in the next time step is the Born-rule probability reweighted by the ratio of the probabilities that it is accepted after outcome $\mu$ and from its current state. To address this question concretely, we use a soft form of post-selection, which we refer to as biasing. Let 
\begin{equation}\label{eq:intro-AT}
    A_T\coloneqq\int_0^Tf\big(\ket{\psi(t)}\big)\,\dd t
\end{equation}
be the time integral, up to a final time $T$, of a function $f$ of the post-measurement state. Biasing multiplies the Born probability of each record by $e^{-sA_T}$, which for $s<0$ favours trajectories along which $f$ is large and for $s>0$ those along which it is small; when $e^{-sA_T}\le1$ this is equivalent to accepting each trajectory with probability $e^{-sA_T}$. Such Bayesian ideas have been applied to the post-measurement state of small systems by Carollo, Garrahan and Jack~\cite{CGJ21} with a nonlinear protocol. 

In this work, we extend such Bayesian ideas to the post-selection of nonlinear properties, such as the von Neumann entropy, in many-body systems and establish their physical realisation as quantum instruments. We show that, whenever the reachable post-measurement states split into finitely many sets on each of which $f$ is constant, the biased trajectories are produced without post-selection by a quantum instrument that acts on the system and on an additional register. We refer to it as the optimal controlled quantum instrument (OCQI), as its rates follow from a variational principle~\cite{CGJ21} (Sec.~\ref{sec:optimal}): per unit time, they minimise the Kullback--Leibler divergence of the records from those of the original dynamics plus $s$ times the mean of $A_T$. Furthermore, the OCQI has a natural tensor-network representation, which extends the cMPS of Eq.~\eqref{eq:intro-traj}, generated sequentially by the monitored system~\cite{SSVCW05,VC10,OEV10}. 

Section~\ref{sec:setting} describes monitored dynamics as a quantum instrument, introduces biasing, and derives the optimal dynamics from Bayes' rule,
and Sec.~\ref{sec:instrument} constructs the optimal controlled quantum instrument with tensor networks. We illustrate these results with a single
qubit that precesses in a magnetic field, biased by its spin component along the field (Sec.~\ref{sec:spin}), and with free fermions with gain and
loss on every site, biased by the squared deviation of their particle number from half filling (Sec.~\ref{sec:ff}). In Appendix~\ref{sec:chain}, we bias
a chain of Majoranas whose neighbouring parities are measured~\cite{NS20} by the von Neumann entropy of the post-measurement state and, using the
Temperley--Lieb algebra~\cite{TL71} (Appendix~\ref{app:tl}) and quantum integrability techniques, argue that this leads to a purification transition
under one additional assumption. Section~\ref{sec:concl} summarises the results and discusses open problems. 
\section{Background: Biased quantum dynamics}\label{sec:setting}

\subsection{Monitored dynamics}\label{sec:monitored}

Consider a system with Hilbert space $\mathbb C^d$ coupled to an environment. Averaged over everything that the environment records, the system evolves
under the Lindbladian of Eq.~\eqref{eq:intro-lindblad}, which in terms of the non-Hermitian effective Hamiltonian $\Heff=H-\frac i2\Gamma$, with
$\Gamma\coloneqq\sum_\mu L_\mu^\dagger L_\mu$, reads $\mathcal L(\rho)=-i\big(\Heff\rho-\rho\Heff^\dagger\big)+\sum_\mu L_\mu\rho L_\mu^\dagger$. The anti-Hermitian part of $\Heff$
accounts for the probability carried away by the quantum jumps, since $\braket{\psi|\Gamma|\psi}$ is the total rate at which a jump occurs from $\ket\psi$.

\begin{figure}[t]
\centering
\includegraphics{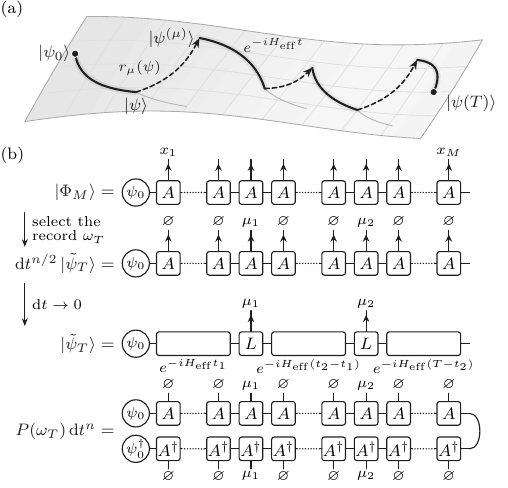}
\caption{The post-measurement state and the MPS of the record. (a)~While the outcome is $\varnothing$, the post-measurement state follows $e^{-i\Heff t}$, normalised (solid lines), and an outcome $\mu$ takes it to $\ket{\psi^{(\mu)}}$ at the rate $r_\mu(\psi)$ (dashed arrows). (b)~The MPS $\ket{\Phi_M}$ of Eq.~\eqref{eq:II-mps} (first row), its component along a record $\omega_T$, which is $\dd t^{n/2}\ket{\tilde\psi_T}$ to leading order in $\dd t$ (second row), the limit $\dd t\to0$, which gives $\ket{\tilde\psi_T}$ of Eq.~\eqref{eq:II-unnorm} (third row), and the Born probability of the record, with its outcomes on the physical legs of both layers (bottom row). Dotted lines stand for sequences of steps.}
\label{fig:monitored}
\end{figure}

Suppose now that the environment is monitored, so that each jump is detected and its type $\mu$ is known. Over a short
interval $\dd t$ the system then undergoes the instrument with the Kraus operators
\begin{equation}\label{eq:kraus}
  K_\varnothing=\one_{\mathrm{sys}}-i\Heff\,\dd t,\qquad K_\mu=\sqrt{\dd t}\,L_\mu ,
\end{equation}
where $\one_{\mathrm{sys}}$ is the identity of the system, the outcome $\mu$ means that a jump of type $\mu$ was detected and the outcome $\varnothing$ that none was. Summed over the outcome, they reproduce
the Lindbladian,
\[
  K_\varnothing\rho K_\varnothing^\dagger+\sum_\mu K_\mu\rho K_\mu^\dagger=\rho+\dd t\,\mathcal L(\rho)+\mathcal O(\dd t^2).
\]
Recording the outcomes up to a final time $T$ yields the quantum-jump unravelling of $\mathcal L$. We divide $[0,T]$ into $M\coloneqq T/\dd t$ steps
of length $\dd t$ and denote the outcome of the $k$th step by $x_k\in\Out$, where the set $\Out$ of outcomes contains $\varnothing$ and one outcome $\mu$
for each jump operator $L_\mu$. With the $d\times d$ Kraus operators of
Eq.~\eqref{eq:kraus} as matrices, $A^\varnothing\coloneqq K_\varnothing$ and $A^\mu\coloneqq K_\mu$, the system and the record are then in the state
\begin{equation}\label{eq:II-mps}
  \ket{\Phi_M}\coloneqq\sum_{x_1,\dots,x_M\in\Out}A^{x_M}\cdots A^{x_1}\ket{\psi_0}\otimes\ket{x_1\cdots x_M}
\end{equation}
in $\mathbb C^d\otimes(\mathbb C^{|\Out|})^{\otimes M}$, see Fig.~\ref{fig:monitored}(b), top row. This is a uniform MPS with open boundary conditions, of
physical dimension $|\Out|$ and bond dimension $d$. Its boundary vector at the first site is $\ket{\psi_0}$, and the auxiliary index of the last
site is left open, since it carries the system at time $T$. The normalisation of the instrument, $\sum_xA^{x\dagger}A^x=\one_{\mathrm{sys}}+\mathcal O(\dd t^2)$,
states that $\ket\phi\mapsto\sum_x\ket x\otimes A^x\ket\phi$ is an isometry up to $\mathcal O(\dd t^2)$, i.e., that the MPS is in the canonical form of a sequentially generated
MPS~\cite{SSVCW05,PVWC07}, so that $\braket{\Phi_M|\Phi_M}=1+\mathcal O(T\,\dd t)$ for every normalised $\ket{\psi_0}$. In the
limit $\dd t\to0$, $\ket{\Phi_M}$ becomes the cMPS of Eq.~\eqref{eq:intro-traj}, of the same bond dimension, whose matrices are $-i\Heff$ and the
$L_\mu$~\cite{VC10}.

The record is the list of the outcomes other than $\varnothing$ and of their times,
\begin{equation}\label{eq:record}
  \omega_T\coloneqq\big((t_1,\mu_1),\dots,(t_n,\mu_n)\big),
\end{equation}
with $0<t_1<\dots<t_n<T$, which labels a basis state of the record in Eq.~\eqref{eq:intro-traj},
$\ket{\omega_T}\coloneqq a_{\mu_n}^\dagger(t_n)\cdots a_{\mu_1}^\dagger(t_1)\ket\varnothing$. In terms of the steps, $x_k=\mu_i$ if the $k$th step contains $t_i$ and $x_k=\varnothing$ otherwise. Expanding the time-ordered exponential, the
component of the cMPS along $\ket{\omega_T}$ is the unnormalised state of the system after the record, see Fig.~\ref{fig:monitored}(b), third row,
\begin{align}
  \ket{\tilde\psi_T}\coloneqq{}&e^{-i\Heff(T-t_n)}L_{\mu_n}e^{-i\Heff(t_n-t_{n-1})}\cdots\label{eq:II-unnorm}\\
  &\times e^{-i\Heff(t_2-t_1)}L_{\mu_1}e^{-i\Heff t_1}\ket{\psi_0},\nonumber
\end{align}
and the component of $\ket{\Phi_M}$ along the record, i.e., along the outcomes $x_k$ that it fixes, see Fig.~\ref{fig:monitored}(b), second row, is,
to leading order in $\dd t$,
\begin{equation}\label{eq:II-Aprod}
  A^{x_M}\cdots A^{x_1}\ket{\psi_0}=\dd t^{\,n/2}\,\ket{\tilde\psi_T}.
\end{equation}
Normalising $\ket{\tilde\psi_T}$ gives the post-measurement state at time $T$,
$\ket{\psi(T)}\coloneqq\ket{\tilde\psi_T}/\braket{\tilde\psi_T|\tilde\psi_T}^{1/2}$. 

For a record in the given basis, quantum theory gives three rules~\cite{WM09,Car93,DCM92,PK98}:
\begin{itemize}
\item[(a)] From $\ket\psi$, the measurement outcome $\mu$ occurs at the rate
\begin{equation}\label{eq:rates}
  r_\mu(\ket\psi)\coloneqq\braket{\psi|L_\mu^\dagger L_\mu|\psi}
\end{equation}
and changes it to the post-measurement state $\ket{\psi^{(\mu)}}\coloneqq L_\mu\ket\psi/\sqrt{r_\mu(\ket\psi)}$.
\item[(b)] While the outcome is $\varnothing$, the post-measurement state evolves deterministically, from $\ket\psi$ to
\begin{equation}\label{eq:II-evol}
  \ket{\psi(t)}=\frac{e^{-i\Heff t}\ket\psi}{\braket{\psi|e^{i\Heff^\dagger t}e^{-i\Heff t}|\psi}^{1/2}}
\end{equation}
after a time $t$.
\item[(c)] The Born probability density of the record, with respect to $\dd t_1\cdots\dd t_n$, is the squared norm of the component of the cMPS along $\ket{\omega_T}$,
\begin{align}
  P(\omega_T)&=\braket{\tilde\psi_T|\tilde\psi_T}\label{eq:pathdensity}\\
  &=\prod_{i=1}^nr_{\mu_i}\big(\ket{\psi(t_i^-)}\big)\,\exp\Big(-\int_0^TR\big(\ket{\psi(t)}\big)\,\dd t\Big),\nonumber
\end{align}
where $R\coloneqq\sum_\mu r_\mu$ is the total rate and $\ket{\psi(t_i^-)}$ denotes the state just before the outcome $\mu_i$ at time $t_i$.
\end{itemize}

Rule (a) is the Born rule for $K_\mu$, $\braket{\psi|K_\mu^\dagger K_\mu|\psi}=r_\mu(\ket\psi)\dd t$, and rule (b) is $K_\varnothing$ applied
repeatedly. Rule (c) holds because while the outcome is $\varnothing$, $\frac{\dd}{\dd t}\braket{\tilde\psi|\tilde\psi}=-\braket{\tilde\psi|\Gamma|\tilde\psi}=-R\big(\ket{\psi(t)}\big)\braket{\tilde\psi|\tilde\psi}$,
each outcome $\mu$ multiplies $\braket{\tilde\psi|\tilde\psi}$ by $r_\mu$, and the factors $\dd t$ carried by $K_\mu=\sqrt{\dd t}\,L_\mu$ are divided out, which makes $P$ a
density in the times $t_i$ of the outcomes. The probability of the record is therefore the squared norm $P(\omega_T)\,\dd t^{\,n}$ of the component of $\ket{\Phi_M}$ in
Eq.~\eqref{eq:II-Aprod}, see Fig.~\ref{fig:monitored}(b), bottom row. 

Similarly, the same rules hold for mixed states with $\rho$ in place of $\ket\psi\bra\psi$, i.e., with rate $\Tr(L_\mu^\dagger L_\mu\rho)$,
post-measurement state $L_\mu\rho L_\mu^\dagger/\Tr(\cdot)$ and evolution $e^{-i\Heff t}\rho\, e^{i\Heff^\dagger t}/\Tr(\cdot)$ for outcome $\varnothing$. The record together with the path $\ket{\psi(t)}$, $0\le t\le T$, is a quantum trajectory, which, given $\ket{\psi_0}$, is specified by its record. By
rules (a) and (b), the future of $\ket{\psi(t)}$ depends on the past record only through $\ket{\psi(t)}$, i.e., the post-measurement state performs a
random walk on pure states, in which a deterministic evolution is interrupted by outcomes $\mu$ at random times, see Fig.~\ref{fig:monitored}(a).

\subsection{Biased trajectories}\label{sec:biasing}

The quantities of interest are often nonlinear functions $f(\ket\psi)$ of the post-measurement state, whose average
$\E\big[f\big(\ket{\psi(T)}\big)\big]$ is not a function of the averaged state $e^{T\mathcal L}(\ket{\psi_0}\bra{\psi_0})$. Such averages are over the reachable
post-measurement states, i.e., the states $\ket{\psi(t)}$ along the records of nonzero probability density started from $\ket{\psi_0}$.

To select trajectories by such a property, we accept or reject each trajectory according to its record. With a real function $f(\ket\psi)$
of the post-measurement state and a real number $s$, a trajectory in the post-measurement state $\ket\psi$ is accepted over each interval $\dd t$
with probability $e^{-s f(\ket\psi)\dd t}$, which requires $sf\ge0$. A trajectory with the record $\omega_T$ is therefore accepted with probability
\begin{equation}\label{eq:bias}
  W(\omega_T)\coloneqq e^{-sA_T},
\end{equation}
with $A_T$ of Eq.~\eqref{eq:intro-AT}, the time integral of $f$ along the trajectory, which we call the bias of the record. We use Eq.~\eqref{eq:bias} for any real $s$ and $f$, in which case $W$ can exceed one. The biased ensemble then
assigns to each record the probability density
\begin{equation}\label{eq:Q}
  Q(\omega_T)\coloneqq\frac{P(\omega_T)\,W(\omega_T)}{Z_T},\qquad Z_T\coloneqq\E[W],
\end{equation}
where the average comprises a sum over the number and types of the outcomes other than $\varnothing$ and an integral over their times $t_i$. Regarding $W$ as an operator on the record, diagonal in the records, $Z_T$ is its expectation in the cMPS of
Eq.~\eqref{eq:intro-traj}. We call $Z_T$ the partition function
of trajectories and, with $\E_\psi$ the average over trajectories started from $\ket\psi$, $Z_T(\ket\psi)\coloneqq\E_\psi[W]$ the partition
function of trajectories started from $\ket\psi$. The partition function grows or decays exponentially in $T$, at the rate
\begin{equation}\label{eq:theta}
  \theta\coloneqq\lim_{T\to\infty}\frac1T\ln Z_T .
\end{equation}

As $Z_T=\E[e^{-sA_T}]$ is the moment generating function of $A_T$, $\partial_s\ln Z_T=-\E_Q[A_T]$, with $\E_Q$ the average in the biased
ensemble, so that the mean of $A_T/T$ in the biased ensemble is
\begin{equation}\label{eq:II-mean}
  -\theta'(s)=\lim_{T\to\infty}\frac{\E_Q[A_T]}T,
\end{equation}
which at $s=0$ is the typical value of $A_T/T$. Other values $\bar a$ are exponentially unlikely at long times, known as the large deviation principle~\cite{DV75,DZ98,Tou09,LAvW07,JS10,GL10},
\begin{equation}\label{eq:II-ldp}
  \mathrm{Prob}\big[A_T/T\approx\bar a\big]\asymp e^{-TI(\bar a)},
\end{equation}
where $I(\bar a)=\sup_s\big(-s\bar a-\theta(s)\big)$, and $\asymp$ means that the logarithms agree to leading order in $T$.

\subsection{Optimal dynamics from Bayes' rule}\label{sec:optimal}

When $W\le1$, post-selection costs about $e^{-T\theta}$ repeated experiments per accepted trajectory. Here we construct an optimal dynamics whose
typical trajectories are the biased ones. At long times, the partition function of trajectories started from a reachable state $\ket\psi$ grows at the
same rate $\theta$, with a prefactor
\begin{equation}\label{eq:g}
  g(\ket\psi)\coloneqq\lim_{T\to\infty}e^{-T\theta}Z_T(\ket\psi),
\end{equation}
which we assume to exist and to be positive. It is in general a
nonlinear function of $\ket\psi$. We consider long trajectories, $T\to\infty$, and times $t$ far from $T$, so that
$Z_{T-t}(\ket\phi)\approx e^{(T-t)\theta}g(\ket\phi)$ for every reachable state $\ket\phi$.

Consider a trajectory of the original dynamics that is in $\ket\psi$ at time $t$ and will be accepted. It is accepted during the remaining time with probability $Z_{T-t}(\ket\psi)$, and after an outcome $\mu$ with probability $Z_{T-t}(\ket{\psi^{(\mu)}})$. By Bayes' rule, the probability of the outcome $\mu$ in the next interval $\dd t$, given that the trajectory will be accepted, is the Born probability $r_\mu\dd t$ multiplied by the ratio of these two probabilities,
\begin{equation*}
\begin{split}
  \mathrm{Prob}\big[\mu\,\big|\,\text{accepted}\big]&=\frac{\mathrm{Prob}\big[\text{accepted}\,\big|\,\mu\big]\,\mathrm{Prob}[\mu]}{\mathrm{Prob}[\text{accepted}]}\\
  &=\frac{Z_{T-t}(\ket{\psi^{(\mu)}})\,r_\mu(\ket\psi)\,\dd t}{Z_{T-t}(\ket\psi)}.
\end{split}
\end{equation*}
For long trajectories, it defines the rate
\begin{align}
  r^g_\mu(\ket\psi)&\coloneqq\lim_{T\to\infty}\frac{\mathrm{Prob}\big[\mu\,\big|\,\text{accepted}\big]}{\dd t}\label{eq:optrates}\\
  &=r_\mu(\ket\psi)\,\frac{g(\ket{\psi^{(\mu)}})}{g(\ket\psi)} ,\nonumber
\end{align}
with $r^g_\mu(\ket\psi)\coloneqq0$ where $r_\mu(\ket\psi)=0$.

The ratio $g(\ket{\psi^{(\mu)}})/g(\ket\psi)$ favours the outcomes $\mu$ that lead to states from which the biased trajectories are easy to continue. Away from the final time $T$, the records of this optimal dynamics are distributed as the biased ensemble $Q$.

In addition, the optimal dynamics is supported by a variational principle~\cite{CGJ21}. Consider all dynamics that keep the evolution for outcome
$\varnothing$ and change the rates of the outcomes $\mu$ to non-negative functions $k_\mu(\ket\psi)$ of the post-measurement state. Among these dynamics, the optimal one minimises, per unit time at long times, the sum of $s$ times the mean of $A_T$ and the Kullback--Leibler divergence
of its records from those of the original dynamics, and the minimum is $-\theta$.

Over an interval $\dd t$, the optimal dynamics applies to $\ket\psi$ the operators
\begin{equation}\label{eq:II-krausopt}
  \big(g(\ket{\psi^{(\mu)}})/g(\ket\psi)\big)^{1/2}K_\mu ,
\end{equation}
for the outcomes $\mu$, and $\big(1-\tfrac12(R^g-R)\,\dd t\big)K_\varnothing$, with $R^g\coloneqq\sum_\mu r^g_\mu$, for the outcome $\varnothing$.
These operators give the outcome $\mu$ with probability $r^g_\mu\dd t$ and the same post-measurement states as the original Kraus operators, while the outcome
$\varnothing$ occurs with the remaining probability, $1-R^g\dd t$, and leaves the evolution for outcome $\varnothing$ unchanged.
\providecommand{\HeffG}{H^{G}_{\mathrm{eff}}}
\providecommand{\HeffBias}{H^{W}_{\mathrm{eff}}}
\providecommand{\LbiasW}{\mathcal{L}_{W}}

\section{The optimal controlled quantum instrument}\label{sec:instrument}

In general, the operators of Eq.~\eqref{eq:II-krausopt} do not form a quantum instrument, because they depend on the post-measurement state through the nonlinear function $g$. In this section we construct a quantum instrument that produces the biased trajectories, with the optimal rates
of Eq.~\eqref{eq:optrates}. 
\subsection{Partitions}\label{sec:register}

\begin{figure}[t]
\centering
\includegraphics{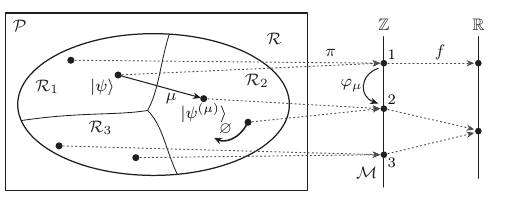}
\caption{A partition for the bias. The reachable states $\Reach$ (ellipse) among the post-measurement states $\Pst$ (rectangle) are split into the sets $\Reach_1$, $\Reach_2$ and $\Reach_3$, the map $\pi$ sends $\Reach_m$ to $m\in\Mem$, and $f$ is a function on $\Mem$. An outcome $\mu$ takes $\Reach_1$ into the single set $\Reach_{\varphi_\mu(1)}=\Reach_2$, and the evolution for outcome $\varnothing$ stays within each set.}
\label{fig:memory}
\end{figure}

The bias of Eq.~\eqref{eq:bias} is determined by the values of $f$ at the post-measurement states along the trajectory. The post-measurement
states are elements of the set $\Pst\coloneqq\big\{\ket\psi\in\mathbb C^d:\ \braket{\psi|\psi}=1\big\}/\mathrm U(1)$ of unit vectors up to a
global phase, and $f:\Pst\to\mathbb R$ is a real function.

\begin{definition}[reachable states]\label{def:reach}
The set of reachable states is the smallest set $\Reach\subseteq\Pst$ that contains $\ket{\psi_0}$ and, with each $\ket\psi\in\Reach$, contains
$\ket{\psi(t)}$ for all $t\ge0$ and $\ket{\psi^{(\mu)}}$ for all $\mu\in\Out\setminus\{\varnothing\}$ with $r_\mu(\ket\psi)>0$.
\end{definition}

\noindent The set $\Reach$ contains the post-measurement states $\ket{\psi(T)}$ after all records $\omega_T$ of nonzero probability density, for all
$T\ge0$, together with the states $\ket{\psi^{(\mu)}}$ right after each outcome $\mu$. In general, the set is infinite because the evolution for outcome
$\varnothing$ of Eq.~\eqref{eq:II-evol} changes a state that is not an eigenvector of $\Heff$ continuously in time.

\begin{definition}[partition for the bias]\label{def:partition}
Let $\Mem$ be a set, and let $\Reach=\dunion_{m\in\Mem}\Reach_m$ be a partition of the reachable states into nonempty, pairwise disjoint sets
$\Reach_m$. We call it a partition for the bias if there are maps $\varphi_\mu:\Mem\to\Mem$, one for each $\mu\in\Out\setminus\{\varnothing\}$, such that
for all $m\in\Mem$
\begin{itemize}
\item[(a)] $f$ is constant on $\Reach_m$;
\item[(b)] $\ket{\psi(t)}\in\Reach_m$ for all $\ket\psi\in\Reach_m$ and $t\ge0$;
\item[(c)] $\ket{\psi^{(\mu)}}\in\Reach_{\varphi_\mu(m)}$ for all $\ket\psi\in\Reach_m$ and $\mu\in\Out\setminus\{\varnothing\}$ with
$r_\mu(\ket\psi)>0$.
\end{itemize}
\end{definition}

\noindent Note that condition (a) depends on $f$, while conditions (b) and (c) do not involve $f$. By (a) and (b), $f$ must be constant along the
evolution for outcome $\varnothing$ of every reachable state, so that not every $f$ has a partition for the bias. We consider only $f$ that satisfy the following assumption.
\begin{assumption}\label{ass:finite}
The function $f$ has a partition for the bias $\Reach=\dunion_{m\in\Mem}\Reach_m$ with a finite set $\Mem$.
\end{assumption}

\noindent Under this assumption, we fix one such partition and number its sets, so that $\Mem\subseteq\mathbb Z$.

For any partition for the bias $\Reach=\dunion_{m\in\Mem}\Reach_m$, each $\ket\psi\in\Reach$ lies in exactly one set $\Reach_m$, and we
write $\pi(\ket\psi)$ for this $m$, such that $\pi:\Reach\to\Mem$ is surjective. We write $f(m)$ for the value of $f$ on $\Reach_m$, such that
$f(\ket\psi)=f\big(\pi(\ket\psi)\big)$, see Fig.~\ref{fig:memory}. Along a trajectory, we write $m(t)\coloneqq\pi\big(\ket{\psi(t)}\big)$ and
$m_0\coloneqq\pi(\ket{\psi_0})$.

\begin{lemma}\label{lem:partition}
Let $\Reach=\dunion_{m\in\Mem}\Reach_m$ be a partition for the bias.
\begin{itemize}
\item[(i)] Along every record of nonzero probability density, with the outcomes $\mu_1,\dots,\mu_n$ at the times $t_1<\dots<t_n$, $m(t)$ is constant
between consecutive outcomes, equal to $m_i$ for $t_i<t<t_{i+1}$, where $t_0\coloneqq0$ and $t_{n+1}\coloneqq T$, and
$m_i=\varphi_{\mu_i}(m_{i-1})$ for $i=1,\dots,n$. Hence $m(t)$ is determined by $m_0$ and the record through the maps $\varphi_\mu$.
\item[(ii)] The bias of Eq.~\eqref{eq:bias} is
\begin{equation*}
  W(\omega_T)=\exp\Big(-s\int_0^Tf\big(m(t)\big)\,\dd t\Big).
\end{equation*}
\item[(iii)] For all $m\in\Mem$ and $\mu\in\Out\setminus\{\varnothing\}$, $\Heff\,\operatorname{span}\Reach_m\subseteq\operatorname{span}\Reach_m$ and
$L_\mu\operatorname{span}\Reach_m\subseteq\operatorname{span}\Reach_{\varphi_\mu(m)}$, with $L_\mu\operatorname{span}\Reach_m=0$ if $r_\mu$ vanishes on $\Reach_m$.
\end{itemize}
\end{lemma}

\begin{proof}
(i) By (b), the evolution for outcome $\varnothing$ keeps the post-measurement state in one set $\Reach_m$ between consecutive outcomes, so
that $m(t)$ is constant there, with $m(t)=m_0$ for $t<t_1$. The outcome $\mu_i$ occurs from the state reached just before it, which lies in
$\Reach_{m_{i-1}}$ by (b) and has $r_{\mu_i}>0$, since the record has nonzero probability density. By (c), it leads to a state in
$\Reach_{\varphi_{\mu_i}(m_{i-1})}$, so that $m_i=\varphi_{\mu_i}(m_{i-1})$.

(ii) By (a), $f\big(\ket{\psi(t)}\big)=f\big(m(t)\big)$ for all $t$, which inserted in Eq.~\eqref{eq:bias} gives (ii).

(iii) For $\ket\psi\in\Reach_m$ and $t\ge0$, $e^{-i\Heff t}\ket\psi$ is proportional to
$\ket{\psi(t)}$, which lies in $\Reach_m$ by (b). Hence $e^{-i\Heff t}\ket\psi\in\operatorname{span}\Reach_m$, and so is its derivative $-i\Heff\ket\psi$ at
$t=0$. For $\ket\psi\in\Reach_m$, $L_\mu\ket\psi$ is proportional to $\ket{\psi^{(\mu)}}\in\Reach_{\varphi_\mu(m)}$ if $r_\mu(\ket\psi)>0$, by (c),
and vanishes if $r_\mu(\ket\psi)=\|L_\mu\ket\psi\|^2=0$. The statements on $\operatorname{span}\Reach_m$ follow by linearity.
\end{proof}

\noindent By (i) and (ii), computing the bias requires $m(t)$, which takes values in the finite set $\Mem$, but never the post-measurement state itself.

To each $m\in\Mem$ we attach a subspace $\mathcal S_m\subseteq\mathbb C^d$ of the system that contains $\Reach_m$ and on which the dynamics
acts as on $\Reach_m$, i.e.,
\begin{equation}\label{eq:Sm}
  \Reach_m\subseteq\mathcal S_m,\quad\Heff\mathcal S_m\subseteq\mathcal S_m,\quad L_\mu\mathcal S_m\subseteq\mathcal S_{\varphi_\mu(m)}
\end{equation}
for all $m\in\Mem$ and $\mu\in\Out\setminus\{\varnothing\}$, so that the evolution for outcome $\varnothing$ stays in $\mathcal S_m$ and the
outcome $\mu$ takes $\mathcal S_m$ into $\mathcal S_{\varphi_\mu(m)}$. We write $P_m$ for the orthogonal projector onto $\mathcal S_m$. By
Lemma~\ref{lem:partition}(iii), $\mathcal S_m=\operatorname{span}\Reach_m$ satisfies Eq.~\eqref{eq:Sm}, and every other choice contains it. The subspaces $\mathcal S_m$ need not be mutually orthogonal, and they
cannot be if a state of $\Reach_m$ overlaps with a state of $\Reach_{m'}$ for some $m'\ne m$.

\subsection{The bias MPS}\label{sec:biasmps}

\begin{figure}[t]
\centering
\includegraphics{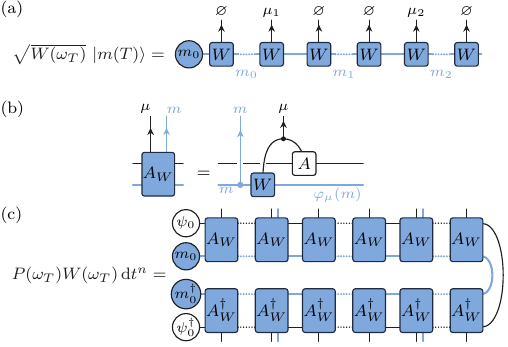}
\caption{The bias MPS and its product with the MPS of the record, for a record like that of Fig.~\ref{fig:monitored}(b). (a)~The bias MPS, with the matrices $W^x$ of Eq.~\eqref{eq:III-Wmps} and the register as auxiliary space (blue), follows $m(t)$ and gives $\sqrt{W(\omega_T)}\,\ket{m(T)}$ to leading order in $\dd t$. (b)~One step of the product, for an outcome $\mu$: the GHZ tensor (black dot) makes $A$ and $W$ read the same outcome, and a GHZ tensor (blue dot), which gives $A_W^{(\mu,m)}$; for the outcome $\varnothing$, $A_W$ has the black leg only. (c)~The double layer of the product gives $P(\omega_T)W(\omega_T)\,\dd t^n$ to leading order in $\dd t$, with each outcome $\mu_i$ recorded together with $m_{i-1}$ (blue legs).}
\label{fig:biasmps}
\end{figure}

We consider a register with the orthonormal basis states $\ket m$, $m\in\Mem$, which stores $m$ and holds $\ket{m(t)}$ at time $t$. The bias is then given by an MPS of the form of Eq.~\eqref{eq:II-mps}, the bias MPS, with the sites and the physical index of $\ket{\Phi_M}$ and the register as auxiliary space: the boundary vector $\ket{m_0}$ replaces $\ket{\psi_0}$, and the matrices $W^x$ replace $A^x$. The auxiliary index takes the values $m\in\Mem$, so that the bond
dimension is $|\Mem|$, and the $|\Mem|\times|\Mem|$ matrices $W^x$ have the elements
\begin{align}
  \braket{m'|W^\varnothing|m}&\coloneqq e^{-sf(m)\dd t/2}\,\delta_{m'm},\label{eq:III-Wmps}\\
  \braket{m'|W^\mu|m}&\coloneqq\delta_{m',\varphi_\mu(m)},\nonumber
\end{align}
for $m,m'\in\Mem$. Here and below, identities between the matrices of one step hold to first order in $\dd t$, and those for their products over a time $T$ to leading order in $\dd t$, which is all that the continuous-time results, obtained in the limit $\dd t\to0$, require.
By Lemma~\ref{lem:partition}(i) and (ii), the auxiliary index follows $m(t)$ along a record of nonzero probability density, and the component of the bias MPS along the record is, to leading order in $\dd t$, $W^{x_M}\cdots W^{x_1}\ket{m_0}=\sqrt{W(\omega_T)}\,\ket{m(T)}$,
and its squared norm is the bias, see Fig.~\ref{fig:biasmps}(a),
\begin{equation}\label{eq:III-Wnorm}
  W(\omega_T)=\bra{m_0}W^{x_1\dagger}\cdots W^{x_M\dagger}\,W^{x_M}\cdots W^{x_1}\ket{m_0},
\end{equation}
just as the squared norm of the component of $\ket{\Phi_M}$ is $P(\omega_T)\,\dd t^{\,n}$. By Assumption~\ref{ass:finite}, the bias is thus an MPS of finite bond dimension
$|\Mem|$ over the record.

\subsection{Dilation via a register}\label{sec:product}
The optimal rates of Eq.~\eqref{eq:optrates} are in general functions of the whole post-measurement state and not of $m$ alone. The system and the register must share the same record, and we therefore combine $\ket{\Phi_M}$ and the bias MPS into a single
MPS, $\ket{\Phi_M^W}$. It is obtained by contracting the physical legs of $A$ and $W$ with the GHZ tensor $\sum_x\ket x\bra x\bra x$ for every
site, such that both share the same outcome $x$, see Fig.~\ref{fig:biasmps}(b), and by recording each outcome $\mu$ together with the value
$m$ of the register that it overwrites. This value, $m_{i-1}$ for the outcome $\mu_i$, can be inferred from $m_0$ and the outcomes
$\mu_1,\dots,\mu_{i-1}$ by Lemma~\ref{lem:partition}(i), so that recording it requires no access to the register and adds no information to
the record. The physical index thus takes the values $\varnothing$ and the pairs $(\mu,m)$, with the matrices
$A_W^\varnothing\coloneqq A^\varnothing\otimes W^\varnothing$ and $A_W^{(\mu,m)}\coloneqq A^\mu\otimes W^\mu\ket m\bra m$, and the
boundary vector $\ket{\psi_0}\ket{m_0}$. Explicitly, $A_W^\varnothing=\one_{\mathrm{sys}}\otimes\one_{\mathrm{reg}}-i\HeffBias\,\dd t+\mathcal O(\dd t^2)$
and $A_W^{(\mu,m)}=\sqrt{\dd t}\,L^W_{\mu,m}$, with
\begin{align}
  \HeffBias&\coloneqq\Heff\otimes\one_{\mathrm{reg}}-\tfrac i2\,s\,\one_{\mathrm{sys}}\otimes\hat f,\label{eq:III-joint}\\
  L^W_{\mu,m}&\coloneqq L_\mu\otimes\ket{\varphi_\mu(m)}\bra m ,\nonumber
\end{align}
where $\hat f\coloneqq\sum_mf(m)\ket m\bra m$, and $\one_{\mathrm{reg}}$ is the identity of the register. Hence $\ket{\Phi_M^W}$ has system $\otimes$ register as auxiliary space. Along a record of nonzero probability density,
the register is in $\ket{m_{i-1}}$ just before the outcome $\mu_i$, so that only the pair $(\mu_i,m_{i-1})$ contributes, and the components along
all other records of pairs vanish. With these pairs, the component along the record is, to leading order in $\dd t$,
\begin{equation*}
  A_W^{x_M}\cdots A_W^{x_1}\ket{\psi_0}\ket{m_0}=\dd t^{\,n/2}\sqrt{W(\omega_T)}\,\ket{\tilde\psi_T}\ket{m(T)},
\end{equation*}
whose squared norm $P(\omega_T)W(\omega_T)\,\dd t^{\,n}$ is the Born probability of the record times its bias, see Fig.~\ref{fig:biasmps}(c). In
the limit $\dd t\to0$, $\ket{\Phi_M^W}$ becomes the cMPS
\begin{multline}\label{eq:III-prod}
  \mathcal T\exp\Big(\int_0^T\!\dd t\,\Big(-i\HeffBias\otimes\one_{\mathrm{rec}}+\sum_{\mu,m}L^W_{\mu,m}\otimes a_{\mu,m}^\dagger(t)\Big)\Big)\\
  \times\ket{\psi_0}\ket{m_0}\ket\varnothing,
\end{multline}
where $a_{\mu,m}^\dagger(t)$ adds the pair $(\mu,m)$ at time $t$ to the record, and $\one_{\mathrm{rec}}$ is the identity of the record of the pairs.

The MPS $\ket{\Phi_M^W}$ only lives in part of system $\otimes$ register, namely the space
\begin{equation}\label{eq:III-V}
  \mathcal V\coloneqq\bigoplus_m\mathcal S_m\otimes\ket m .
\end{equation}
Its boundary vector lies in $\mathcal V$, because $\ket{\psi_0}\in\mathcal S_{m_0}$, and by Eq.~\eqref{eq:Sm}, $A_W^\varnothing$ maps each
$\mathcal S_m\otimes\ket m$ into itself, while $A_W^{(\mu,m)}$ maps it into $\mathcal S_{\varphi_\mu(m)}\otimes\ket{\varphi_\mu(m)}$ and
annihilates $\mathcal S_{m'}\otimes\ket{m'}$ for $m'\ne m$. Hence
$\ket{\Phi_M^W}$ is an MPS with auxiliary space $\mathcal V$. Note that the direct sum is orthogonal even when the subspaces $\mathcal S_m$ of
the system are not, because the basis states $\ket m$ of the register are orthogonal. The bond dimension is therefore
$\dim\mathcal V=\sum_m\dim\mathcal S_m\le d\,|\Mem|$, and we write $\one_{\mathcal V}=\sum_mP_m\otimes\ket m\bra m$ for the identity on $\mathcal V$.

Summed over records, the squared norm of $\ket{\Phi_M^W}$ is the partition function $Z_T$ of Eq.~\eqref{eq:Q}. We compute it with the
transfer map $\mathcal E_W(\rho)\coloneqq\sum_xA_W^x\rho A_W^{x\dagger}$, where $x$ takes the values $\varnothing$ and $(\mu,m)$, a CP map on the
operators on $\mathcal V$. It equals $e^{\dd t\LbiasW}(\rho)+\mathcal O(\dd t^2)$, where, for operators $\rho$ on $\mathcal V$, and with
$\HeffBias$ and the $L^W_{\mu,m}$ as maps from $\mathcal V$ to itself,
\begin{equation}\label{eq:III-LW}
  \LbiasW(\rho)\coloneqq-i\big(\HeffBias\rho-\rho\,\HeffBias{}^\dagger\big)+\sum_{\mu,m}L^W_{\mu,m}\,\rho\,L^{W\dagger}_{\mu,m} .
\end{equation}
For the initial state $\rho_0\coloneqq\ket{\psi_0}\bra{\psi_0}\otimes\ket{m_0}\bra{m_0}$, $\braket{\Phi_M^W|\Phi_M^W}=\Tr\mathcal E_W^M(\rho_0)$,
so that $Z_T=\Tr\big(e^{T\LbiasW}(\rho_0)\big)$. As $\sum_{\mu,m}L^{W\dagger}_{\mu,m}L^W_{\mu,m}=\sum_mP_m\Gamma P_m\otimes\ket m\bra m$ on $\mathcal V$,
\begin{equation*}
  \LbiasW^\dagger(\one_{\mathcal V})=-s\sum_mf(m)P_m\otimes\ket m\bra m ,
\end{equation*}
so that, for $s=0$, $\LbiasW$ is a Lindbladian, that of the original dynamics with the uncoupled register holding $m(t)$, see Fig.~\ref{fig:tn}(a).

\subsection{The left canonical form}\label{sec:thm}

\begin{figure*}[t]
\centering
\includegraphics{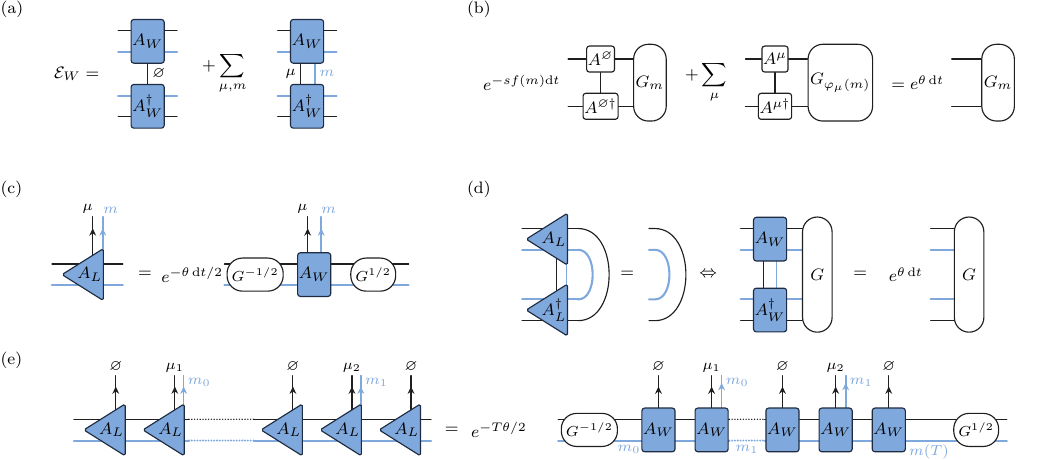}
\caption{The optimal instrument in tensor-network form, over one step $\dd t$, with time running from left to right. Register bonds are blue, and the bra layer is the mirror image of the ket layer. (a)~The transfer map $\mathcal E_W$. The blue leg carries the value $m$ of the register that the outcome $\mu$ overwrites, which is known from the record. (b)~The block $m$ of $\mathcal E_W^\dagger(G)=e^{\theta\,\dd t}G$, an equation for operators on $\mathcal S_m$, i.e., Eq.~\eqref{eq:Geig}. (c)~The gauge transformation by $G$ gives the Kraus operators $A_L^x$ (triangles), shown for $x=(\mu,m)$; for $x=\varnothing$ the blue leg is absent, (d)~which are normalised if and only if $\mathcal E_W^\dagger(G)=e^{\theta\,\dd t}G$, with the joined legs summed over $x$. (e)~Along a record, with the outcomes $\mu_1$ and $\mu_2$ recorded together with $m_0$ and $m_1$ (blue legs), the factors $G^{\pm1/2}$ cancel between neighbouring steps and remain only at the two ends, and the register holds $m(t)$.}
\label{fig:tn}
\end{figure*}

We assume that $\LbiasW$ generates irreducible CP maps with leading eigenvalue $\theta$ and leading left and right eigenvectors $G$ and $\rho_r$,
$\LbiasW^\dagger(G)=\theta G$ and $\LbiasW(\rho_r)=\theta\rho_r$, so that $\ket{\Phi_M^W}$ is normal on $\mathcal V$, i.e., its transfer map is primitive~\cite{CPSV21}, as an irreducible CP semigroup is
primitive~\cite{Wol12}. By the
Perron--Frobenius theorem for CP maps~\cite{EHK78,Wol12}, $\theta$ is real and simple, and $G$ and $\rho_r$ are positive definite on $\mathcal V$.
The eigenvalue $\theta$ is the growth rate of Eq.~\eqref{eq:theta}, since $\Tr\big(G\,e^{T\LbiasW}(\rho_0)\big)=e^{T\theta}\Tr(G\rho_0)$ lies between
the smallest and the largest eigenvalue of $G$ on $\mathcal V$ times $Z_T=\Tr e^{T\LbiasW}(\rho_0)$.
By Lemma~\ref{lem:partition}(iii), $\HeffBias$ and all $L^W_{\mu,m}$ map the subspace $\bigoplus_m\operatorname{span}\Reach_m\otimes\ket m$ of
$\mathcal V$ into itself, so that this assumption can hold only if $\mathcal S_m=\operatorname{span}\Reach_m$ for all $m$.

The leading left eigenvector is block diagonal in the register,
\begin{equation*}
  G=\sum_mG_m\otimes\ket m\bra m ,
\end{equation*}
with $G_m$ positive definite on $\mathcal S_m$. Indeed, as $\HeffBias$ is block diagonal, and each $L^{W\dagger}_{\mu,m}XL^W_{\mu,m}$ is block
diagonal and involves only the diagonal blocks of $X$, the block-diagonal part of $G$ also solves $\LbiasW^\dagger(X)=\theta X$, and as $\theta$ is
simple, it equals $G$. For $s=0$, $\LbiasW^\dagger(\one_{\mathcal V})=0$ and $G=\one_{\mathcal V}$.
In the block $m$, $\LbiasW^\dagger(G)=\theta G$ reads
\begin{align}
  P_m\Big(i[H,G_m]-\tfrac12\{\Gamma,G_m\}&+\sum_\mu L_\mu^\dagger G_{\varphi_\mu(m)}L_\mu\Big)P_m\label{eq:Geig}\\
  &-sf(m)\,G_m=\theta\,G_m ,\nonumber
\end{align}
an equation for operators on $\mathcal S_m$, where the projectors $P_m$ are needed because $H$, $\Gamma$ and
$L_\mu^\dagger G_{\varphi_\mu(m)}L_\mu$ need not map $\mathcal S_m$ into itself. It is a linear problem for $|\Mem|$ operators on the system, coupled through the update of $m$, see Fig.~\ref{fig:tn}(b).

A normal MPS is brought to left canonical form by a gauge transformation of its auxiliary space with the square root of the positive left
eigenvector of its transfer map~\cite{PVWC07,CPSV21}. For $\ket{\Phi_M^W}$ this eigenvector is $G$, and the transformed matrices
\begin{equation}\label{eq:III-gauge}
  A_L^x\coloneqq e^{-\theta\,\dd t/2}\,G^{1/2}A_W^x\,G^{-1/2},
\end{equation}
for $x=\varnothing$ and $x=(\mu,m)$, with $G^{-1/2}$ the inverse of $G^{1/2}$ on $\mathcal V$, are normalised, $\sum_xA_L^{x\dagger}A_L^x=\one_{\mathcal V}$, i.e., their transfer map is TP, with the left
fixed point $\one_{\mathcal V}$, if and only if $\mathcal E_W^\dagger(G)=e^{\theta\,\dd t}G$, see Figs.~\ref{fig:tn}(c) and~\ref{fig:tn}(d). The left canonical form has the boundary vector $G^{1/2}\ket{\psi_0}\ket{m_0}$, normalised, so that its first factor $G^{-1/2}$
gives back $\ket{\psi_0}\ket{m_0}$. Along a record, the factors $G^{-1/2}$ and $G^{1/2}$ of neighbouring steps cancel, and the component
of the left canonical form is $e^{-T\theta/2}G^{1/2}$ times that of $\ket{\Phi_M^W}$, divided by the norm of $G^{1/2}\ket{\psi_0}\ket{m_0}$, see Fig.~\ref{fig:tn}(e). The matrices $A_L^x$ are the Kraus operators of the optimal controlled quantum instrument.

\subsection{The optimal instrument}\label{sec:optimal-instr}

\begin{figure}[t]
\centering
\includegraphics{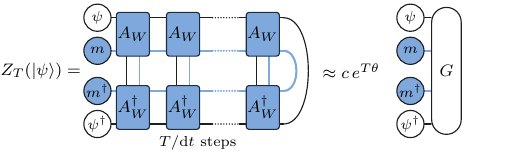}
\caption{The long-time partition function. Starting from $\ket\psi\in\Reach_m$, $Z_T(\ket\psi)$ is the double layer of the product MPS closed by the trace, with the legs of each step joined and summed over $x$, as in Fig.~\ref{fig:tn}(d), which at long times becomes $c\,e^{T\theta}\braket{\psi|G_m|\psi}$ with a constant $c$.}
\label{fig:gquad}
\end{figure}

\begin{theorem}[Optimal Instrument]\label{thm:instrument}
Under Assumption~\ref{ass:finite}, let $\LbiasW$ generate irreducible CP maps on the operators on $\mathcal V$ of Eq.~\eqref{eq:III-V}, with the leading eigenvalue $\theta$
and the leading left eigenvector $G=\sum_mG_m\otimes\ket m\bra m$, and with $G_m^{-1/2}$ the inverse of $G_m^{1/2}$ on $\mathcal S_m$. On $\mathcal V$ let 
\begin{align}
  L^G_{\mu,m}&=G_{\varphi_\mu(m)}^{1/2}L_\mu G_m^{-1/2}\otimes\ket{\varphi_\mu(m)}\bra m,\label{eq:instrument}\\
  \HeffG&=\sum_m\big(G_m^{1/2}\Heff G_m^{-1/2}-\tfrac i2(\theta+sf(m))\big)\otimes\ket m\bra m,\nonumber
\end{align}
which map $\mathcal V$ into itself by Eq.~\eqref{eq:Sm}.
\begin{itemize}
\item[(i)] $\HeffG$ and the $L^G_{\mu,m}$ are the effective Hamiltonian and the jump operators of a Lindbladian on $\mathcal V$, i.e.,
$\HeffG-\HeffG{}^\dagger=-i\sum_{\mu,m}L^{G\dagger}_{\mu,m}L^G_{\mu,m}$, with one jump operator for each pair $(\mu,m)$.
\item[(ii)] Started in $G^{1/2}\ket{\psi_0}\ket{m_0}$, normalised, i.e., in $\ket{\Psi_0}\otimes\ket{m_0}$ with
$\ket{\Psi_0}\coloneqq G_{m_0}^{1/2}\ket{\psi_0}/\braket{\psi_0|G_{m_0}|\psi_0}^{1/2}$, and with each pair $(\mu,m)$ recorded as $\mu$, its
quantum-jump unravelling produces records with the law
\begin{equation}\label{eq:law}
  Q^G(\omega_T)=P(\omega_T)\,W(\omega_T)\,e^{-T\theta}\,\frac{g\big(\ket{\psi(T)}\big)}{g(\ket{\psi_0})},
\end{equation}
where $g(\ket\psi)=\braket{\psi|G_{\pi(\ket\psi)}|\psi}$, with $G$ normalised such that this is $g$ of Eq.~\eqref{eq:g}, which is possible by (iii); Eq.~\eqref{eq:law} does not depend on this normalisation. If the original dynamics is in
$\ket\psi\in\Reach_m$ after a record, the instrument is in $G_m^{1/2}\ket\psi\otimes\ket m$, normalised, i.e., its register is in the basis state $\ket m$.
\item[(iii)] $\braket{\psi|G_{\pi(\ket\psi)}|\psi}$ is, up to a positive factor independent of $\ket\psi$, $g(\ket\psi)$ of Eq.~\eqref{eq:g}, and the outcomes $\mu$ occur at the optimal rates of Eq.~\eqref{eq:optrates}.
\end{itemize}
\end{theorem}

\begin{proof}
By assumption, $\LbiasW^\dagger(G)=\theta G$, i.e., $\mathcal E_W^\dagger(G)=e^{\theta\,\dd t}G$ to first order in $\dd t$.

(i) To first order in $\dd t$, the matrices of Eq.~\eqref{eq:III-gauge} are $A_L^\varnothing=\one_{\mathcal V}-i\HeffG\,\dd t$
and $A_L^{(\mu,m)}=\sqrt{\dd t}\,L^G_{\mu,m}$, so that, to first order, $\sum_xA_L^{x\dagger}A_L^x-\one_{\mathcal V}$ is $\dd t$ times
$i\big(\HeffG{}^\dagger-\HeffG\big)+\sum_{\mu,m}L^{G\dagger}_{\mu,m}L^G_{\mu,m}$. By Sec.~\ref{sec:thm}, it vanishes if and only if $\mathcal E_W^\dagger(G)=e^{\theta\,\dd t}G$. Hence the identity of (i) is equivalent to
$\LbiasW^\dagger(G)=\theta G$, and the $A_L^x$ are the Kraus operators of an instrument, generated by the Lindbladian with the Hamiltonian
$\frac12(\HeffG+\HeffG{}^\dagger)$ and the jump operators $L^G_{\mu,m}$. As $G$ and $\HeffBias$ are block diagonal in the register, these
operators take the block forms of Eq.~\eqref{eq:instrument}.

(ii) The probability density of a record is $\dd t^{-n}$ times the squared norm of the component of the left canonical form along it, with each
outcome $\mu_i$ paired with $m_{i-1}$, the only value of $m$ that contributes (Sec.~\ref{sec:product}), and with the boundary
vector $G^{1/2}\ket{\psi_0}\ket{m_0}/g(\ket{\psi_0})^{1/2}$. By the cancellation of the gauge, this squared norm is $e^{-T\theta}/g(\ket{\psi_0})$
times the trace of $G$ with
\begin{equation*}
\begin{split}
  &A_W^{x_M}\cdots A_W^{x_1}\rho_0\,A_W^{x_1\dagger}\cdots A_W^{x_M\dagger}\\
  &\qquad=\dd t^{\,n}\,W(\omega_T)\,\ket{\tilde\psi_T}\bra{\tilde\psi_T}\otimes\ket{m(T)}\bra{m(T)},
\end{split}
\end{equation*}
which involves only the block $G_{m(T)}$ of $G$. As $\braket{\tilde\psi_T|G_{m(T)}|\tilde\psi_T}=P(\omega_T)\,g\big(\ket{\psi(T)}\big)$, the squared
norm divided by $\dd t^{\,n}$ is Eq.~\eqref{eq:law}. Summed over records, it gives
$e^{-T\theta}\Tr\big(G\,\mathcal E_W^M(\rho_0)\big)/g(\ket{\psi_0})=1$ to leading order in $\dd t$, since $\mathcal E_W^\dagger(G)=e^{\theta\,\dd t}G$ to first order.
By the same cancellation, the instrument is in $G^{1/2}\ket\psi\ket m=G_m^{1/2}\ket\psi\otimes\ket m$, normalised.

(iii) For a reachable state $\ket\psi\in\Reach_m$, $Z_T(\ket\psi)=\Tr e^{T\LbiasW}(\rho)$ with $\rho\coloneqq\ket\psi\bra\psi\otimes\ket m\bra m$, see Fig.~\ref{fig:gquad}. As $\ket{\Phi_M^W}$ is normal, $e^{T\LbiasW}(\rho)\approx e^{T\theta}\rho_r\Tr(G\rho)/\Tr(G\rho_r)$ at long
times~\cite{Wol12}, so that $Z_T(\ket\psi)$ is $e^{T\theta}\braket{\psi|G_m|\psi}$ times a constant, and $\braket{\psi|G_m|\psi}$ is $g(\ket\psi)$ of
Eq.~\eqref{eq:g} up to a factor. By (ii), the outcome $\mu$ occurs in $G_m^{1/2}\ket\psi\ket m/g(\ket\psi)^{1/2}$, at the rate
$\Tr\big(G\,L^W_{\mu,m}\rho L^{W\dagger}_{\mu,m}\big)/g(\ket\psi)$. As $L^W_{\mu,m}\rho L^{W\dagger}_{\mu,m}=L_\mu\ket\psi\bra\psi L_\mu^\dagger\otimes\ket{\varphi_\mu(m)}\bra{\varphi_\mu(m)}$,
this rate is
\begin{equation*}
  \frac{\braket{\psi|L_\mu^\dagger G_{\varphi_\mu(m)}L_\mu|\psi}}{g(\ket\psi)}=r_\mu(\ket\psi)\,\frac{g\big(\ket{\psi^{(\mu)}}\big)}{g(\ket\psi)},
\end{equation*}
since $\pi(\ket{\psi^{(\mu)}})=\varphi_\mu(m)$ by condition (c), which is the optimal rate of Eq.~\eqref{eq:optrates}.
\end{proof}

\subsection{Special cases}\label{sec:special}
We consider two special cases of Theorem~\ref{thm:instrument}, where the optimal instrument can be simplified. 
\subsubsection{Lumpable Partitions}\label{sec:lumpable}

\begin{definition}[Lumpable Partitions]\label{def:lumpable}
The partition for the bias of Assumption~\ref{ass:finite}, with the subspaces $\mathcal S_m$ of Eq.~\eqref{eq:Sm}, is lumpable if, for all $m,m'\in\Mem$ with $m\ne m'$, there is a number
$q(m\to m')\ge0$ such that
\begin{equation}\label{eq:labelop}
  P_m\Big(\sum_{\substack{\mu\in\Out\setminus\{\varnothing\}\\ \varphi_\mu(m)=m'}}L_\mu^\dagger L_\mu\Big)P_m=q(m\to m')\,P_m .
\end{equation}
\end{definition}

\noindent The outcomes that change $m$ to $m'$ then occur at the same total rate $q(m\to m')$ from every state of $\mathcal S_m$, which is the continuous-time form of the condition under which a partition of the states of a Markov
chain is called lumpable~\cite{KS60}.

\begin{theorem}[Lumpable Optimal Instrument]\label{thm:lumpable}
Let the partition for the bias of Assumption~\ref{ass:finite} be lumpable.
\begin{itemize}
\item[(i)] $m(t)=\pi\big(\ket{\psi(t)}\big)$ performs a random walk on $\Mem$ with
the rates $q(m\to m')$.
\item[(ii)] If this walk is irreducible, $\theta$ and $g$ follow from the $|\Mem|\times|\Mem|$ eigenvalue problem
\begin{gather}
  \bra g\Lambda_s=\theta\bra g,\qquad\Lambda_s\coloneqq\Lambda-s\hat f,\label{eq:label}\\
  \Lambda\coloneqq\sum_{m\ne m'}q(m\to m')\big(\ket{m'}\bra m-\ket m\bra m\big),\nonumber
\end{gather}
on the register, where $\Lambda$ is the rate matrix of the walk, whose probabilities $p(m)$ obey $\frac{\dd}{\dd t}\sum_mp(m)\ket m=\Lambda\sum_mp(m)\ket m$, and $\bra g\coloneqq\sum_mg(m)\bra m$: $\theta$ is the leading eigenvalue of $\Lambda_s$,
i.e., its eigenvalue with the largest real part, which is real and simple, $\bra g$ is its leading left eigenvector, with $g$ positive, and $g(\ket\psi)$ of Eq.~\eqref{eq:g} is, up to a
factor, $g\big(\pi(\ket\psi)\big)$.
\item[(iii)] If the walk is irreducible, $G_m=g(m)P_m$ solves Eq.~\eqref{eq:Geig} on $\mathcal S_m$, i.e.,
$G\coloneqq\sum_mg(m)P_m\otimes\ket m\bra m$ solves $\LbiasW^\dagger(G)=\theta G$, and if $\LbiasW$ is moreover irreducible, this $G$ is
its leading left eigenvector up to a factor. The instrument of Theorem~\ref{thm:instrument} with this $G$ has the jump operators and the effective Hamiltonian
\begin{align}
  L^G_{\mu,m}&=\Big(\frac{g\big(\varphi_\mu(m)\big)}{g(m)}\Big)^{1/2}L_\mu\otimes\ket{\varphi_\mu(m)}\bra m ,\label{eq:label-kraus}\\
  \HeffG&=\sum_m\big(\Heff-\tfrac i2(\theta+sf(m))\big)\otimes\ket m\bra m\nonumber
\end{align}
on $\mathcal V$. Started in $\ket{\psi_0}\ket{m_0}$, with each pair $(\mu,m)$ recorded as $\mu$, its
quantum-jump unravelling keeps the system in the post-measurement state $\ket{\psi(t)}$ of the original dynamics and the
register in $\ket{m(t)}$, produces records with the law
\begin{equation}\label{eq:label-law}
  Q^G(\omega_T)=P(\omega_T)\,W(\omega_T)\,e^{-T\theta}\,\frac{g\big(m(T)\big)}{g(m_0)},
\end{equation}
and has the optimal rates of Eq.~\eqref{eq:optrates}, whether or not $\ket{\Phi_M^W}$ is normal.
\end{itemize}
\end{theorem}

\begin{proof}
(i) From a reachable state $\ket\psi\in\Reach_m$, which lies in $\mathcal S_m$, the outcomes $\mu$ with $\varphi_\mu(m)=m'$ change $m$ to $m'$, and by
Eq.~\eqref{eq:labelop} they occur at the total rate $\braket{\psi|\sum_{\mu:\,\varphi_\mu(m)=m'}L_\mu^\dagger L_\mu|\psi}=q(m\to m')$, whatever
$\ket\psi$ is. As $m(t)$ changes only at these outcomes, it jumps from $m$ to $m'$ at the rate $q(m\to m')$.

(ii) By (i) and condition (a), the bias accumulated from $\ket\psi$ depends only on the walk of $m(t)$ started at $\pi(\ket\psi)$.
Conditioning on the first interval $\dd t$, in which $m$ changes to $m'$ with probability $q(m\to m')\dd t$ and the bias is multiplied by
$e^{-sf(m)\dd t}$, gives $Z_T(\ket\psi)=\sum_{m'}\braket{m'|e^{T\Lambda_s}|\pi(\ket\psi)}$. For an irreducible walk, $\Lambda_s$ is irreducible
with non-negative off-diagonal entries, and the Perron--Frobenius theorem~\cite{BP94} gives a real and simple leading eigenvalue $\theta$ with
positive left and right eigenvectors, so that $Z_T(\ket\psi)\approx c\,e^{T\theta}g\big(\pi(\ket\psi)\big)$ at long times, with $c>0$.

(iii) Restricted to $\mathcal S_m$, which $\Heff$ maps into itself and $L_\mu$ into $\mathcal S_{\varphi_\mu(m)}$ by Eq.~\eqref{eq:Sm}, the terms
$i[H,G_m]-\frac12\{\Gamma,G_m\}=i\big(\Heff^\dagger G_m-G_m\Heff\big)$ of Eq.~\eqref{eq:Geig} become $-g(m)P_m\Gamma P_m$, and
$L_\mu^\dagger G_{\varphi_\mu(m)}L_\mu$ becomes $g(\varphi_\mu(m))P_mL_\mu^\dagger L_\mu P_m$. With $\Gamma=\sum_\mu L_\mu^\dagger L_\mu$, the outcomes
$\mu$ with $\varphi_\mu(m)=m$ cancel, and by Eq.~\eqref{eq:labelop} the others give $\sum_{m'\ne m}q(m\to m')\big(g(m')-g(m)\big)P_m=\bra g\Lambda\ket mP_m$. Hence
Eq.~\eqref{eq:Geig} holds on $\mathcal S_m$ with the $\theta$ and $g$ of Eq.~\eqref{eq:label}, and $G_m$ is positive definite on $\mathcal S_m$ since $g>0$.
If $\LbiasW$ is irreducible, its positive solution of $\LbiasW^\dagger(X)=\theta X$ is unique up to a factor, so that $G$ is its leading left
eigenvector. As $G_m$ is a
multiple of $P_m$, the change of frame acts on the register only, and Theorem~\ref{thm:instrument} gives the operators of
Eq.~\eqref{eq:label-kraus}, and as $\braket{\psi|G_m|\psi}=g(m)$ for every
normalised $\ket\psi\in\mathcal S_m$, Eq.~\eqref{eq:law} becomes Eq.~\eqref{eq:label-law}. On $\mathcal S_m\otimes\ket m$, these
operators act as $L_\mu$ and $\Heff$ up to numbers, so that, normalised, the state of the system follows the post-measurement states
$e^{-i\Heff t}\ket\psi$ and $L_\mu\ket\psi$ of the original dynamics. Normality is not needed, since the proof of Theorem~\ref{thm:instrument} uses, for (i), only $\LbiasW^\dagger(G)=\theta G$, and, for (ii), the probability of a record is
$e^{-T\theta}/g(m_0)$ times the trace of $G$ with the block-diagonal operator after the record, which is
$P(\omega_T)W(\omega_T)\,g(m(T))\,\dd t^{\,n}$ by $\braket{\psi|G_m|\psi}=g(m)$. From $\ket\psi\in\Reach_m$, the outcome $\mu$ occurs at the rate
$r_\mu(\ket\psi)\,g(\varphi_\mu(m))/g(m)$, the optimal rate by (ii) of the present theorem.
Finally, if $\LbiasW$ is irreducible, so is the walk, since for a nonempty set $C\subsetneq\Mem$ that the walk cannot leave, Eq.~\eqref{eq:labelop}
gives $L_\mu P_m=0$ for $m\in C$ and $\varphi_\mu(m)\notin C$, so that $\bigoplus_{m\in C}\mathcal S_m\otimes\ket m$ would be mapped into itself by
$\HeffBias$ and all $L^W_{\mu,m}$.
\end{proof}

\subsubsection{Isometry}\label{sec:absorb}

When the subspaces $\mathcal S_m$ are mutually orthogonal, the system itself holds $m$, and the register can be absorbed into it.

\begin{theorem}[Isometry]\label{thm:absorb}
Under the assumptions of Theorem~\ref{thm:instrument}, let the subspaces $\mathcal S_m$ be mutually orthogonal. Then
$V:\ket\phi\otimes\ket m\mapsto\ket\phi$, for $\ket\phi\in\mathcal S_m$, is an isometry from $\mathcal V$ into $\mathbb C^d$, with
$VV^\dagger=\sum_mP_m$, the projector onto $\bigoplus_m\mathcal S_m$, and conjugation by $V$ maps the instrument of Theorem~\ref{thm:instrument} to an instrument on the system:
\begin{itemize}
\item[(i)] Its jump operators, one for each pair $(\mu,m)$, and effective Hamiltonian,
\begin{align}
  VL^G_{\mu,m}V^\dagger&=G_{\varphi_\mu(m)}^{1/2}L_\mu G_m^{-1/2}P_m,\label{eq:absorb}\\
  V\HeffG V^\dagger&=\sum_m\big(G_m^{1/2}\Heff G_m^{-1/2}-\tfrac i2(\theta+sf(m))\big)P_m,\nonumber
\end{align} generate CPTP maps on the operators on $\bigoplus_m\mathcal S_m$. As they vanish on its orthogonal complement, they form a Lindbladian on
the system, which leaves the operators on that complement unchanged; if $\sum_mP_m=\one_{\mathrm{sys}}$, this complement is $\{0\}$.
\item[(ii)] Started in $\ket{\Psi_0}$, with each pair $(\mu,m)$ recorded as $\mu$, its quantum-jump unravelling produces records with the law of
Eq.~\eqref{eq:law}.
\item[(iii)] For a lumpable partition, with $G_m=g(m)P_m$ as in Theorem~\ref{thm:lumpable}(iii), the effective Hamiltonian is
$\sum_m\big(\Heff-\frac i2(\theta+sf(m))\big)P_m$ and the jump operators are $\big(g(\varphi_\mu(m))/g(m)\big)^{1/2}L_\mu P_m$.
\end{itemize}
\end{theorem}

\begin{proof}
As the $\mathcal S_m$ are mutually orthogonal, $V$ maps an orthonormal basis of $\mathcal V$ to an orthonormal basis of
$\bigoplus_m\mathcal S_m$, so that $V^\dagger V=\one_{\mathcal V}$ and $VV^\dagger=\sum_mP_m$, the identity on
$\bigoplus_m\mathcal S_m$.

(i) Conjugation by $V$ maps the Kraus operators $A^x_L$ of the instrument to $\sum_mP_m-iV\HeffG V^\dagger\dd t$ and
$\sqrt{\dd t}\,VL^G_{\mu,m}V^\dagger$, and their normalisation $\sum_xA^{x\dagger}_LA^x_L=\one_{\mathcal V}$ to
$\sum_xVA^{x\dagger}_LV^\dagger\,VA^x_LV^\dagger=\sum_mP_m$. These are therefore the Kraus operators of an instrument on $\bigoplus_m\mathcal S_m$, and
the operators of Eq.~\eqref{eq:absorb} satisfy the identity of Theorem~\ref{thm:instrument}(i) on $\bigoplus_m\mathcal S_m$, and,
as they vanish on its orthogonal complement, also on $\mathbb C^d$.

(ii) The isometry $V$ maps $\ket{\Psi_0}\ket{m_0}$ to $\ket{\Psi_0}$, and the component along every record to its image under $V$, which has
the same norm, so that the law of the records is that of Eq.~\eqref{eq:law}.

(iii) For $G_m=g(m)P_m$, $G_m^{1/2}\Heff G_m^{-1/2}=\Heff P_m$ and $G_{\varphi_\mu(m)}^{1/2}L_\mu G_m^{-1/2}P_m=\big(g(\varphi_\mu(m))/g(m)\big)^{1/2}L_\mu P_m$
on $\mathcal S_m$, by Eq.~\eqref{eq:Sm}.
\end{proof}

\section{Example: A single spin}\label{sec:spin}

\begin{figure}[t]
\centering
\includegraphics{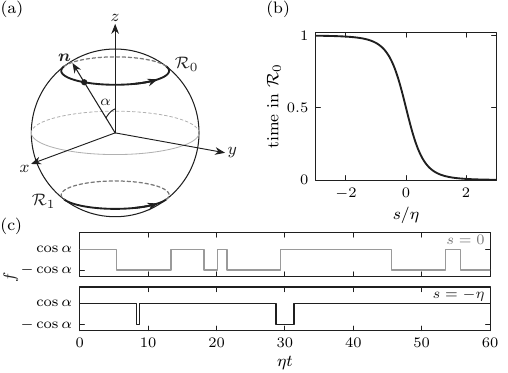}
\caption{A spin $\frac12$ precessing about the field along $z$, measured at the rate $\eta$ along $\bm n$ and biased by $f(\ket\psi)=\braket{\psi|\sigma^z|\psi}$. (a)~The sets $\Reach_0$ and $\Reach_1$ of reachable states on the Bloch sphere, through the post-measurement states $\ket N$ (dot) and $\ket S$ after the outcomes $N$ and $S$; the evolution for outcome $\varnothing$ (arrows) keeps each set. (b)~The fraction of time spent in $\Reach_0$ in the biased ensemble, $\frac12\big(1-\theta'(s)/\cos\alpha\big)$, for $\omega=\eta$ and $\alpha=\pi/3$. (c)~$f$ along a trajectory of the original dynamics ($s=0$) and of the optimal instrument at $s=-\eta$, for the same $\omega$ and $\alpha$.}
\label{fig:spin}
\end{figure}
We illustrate the formal theory with a single spin $\frac12$, which precesses in a magnetic field along the $z$ axis and is measured
projectively, at random times, along an axis at an angle to the field. 
The Hamiltonian is $H=\frac\omega2\sigma^z$, with $\omega\ne0$, and the spin is measured at the rate $\eta$ along $\bm n\coloneqq(\sin\alpha,0,\cos\alpha)$, with
$0<\alpha<\pi/2$, i.e., the jump operators are $L_N\coloneqq\sqrt\eta\,\ket N\bra N$ and $L_S\coloneqq\sqrt\eta\,\ket S\bra S$, where $\ket N$ and
$\ket S$ are the eigenstates of $\bm n\cdot\bm\sigma$, with $\bm\sigma\coloneqq(\sigma^x,\sigma^y,\sigma^z)$, whose Bloch vectors are $\bm n$ and
$-\bm n$. As $\Gamma=\eta\,\one_{\mathrm{sys}}$, the post-measurement state precesses while the outcome is $\varnothing$, which conserves its spin
component along the field, and we bias by this component, $f(\ket\psi)\coloneqq\braket{\psi|\sigma^z|\psi}$. Started in $\ket N$, the reachable states form the set $\Reach_0=\{e^{-iHt}\ket N:t\ge0\}$, on which $f=\cos\alpha$,
and the set $\Reach_1=\{e^{-iHt}\ket S:t\ge0\}$, on which $f=-\cos\alpha$, see Fig.~\ref{fig:spin}(a). They are the two sets of a partition
for the bias (Definition~\ref{def:partition}) with $\Mem=\{0,1\}$, $f(0)=\cos\alpha$ and $f(1)=-\cos\alpha$, in which the outcome $N$ sets the
register to $0$ and the outcome $S$ sets it to $1$, $\varphi_N(0)=\varphi_N(1)=0$ and $\varphi_S(0)=\varphi_S(1)=1$, i.e., the register holds the
latest outcome.

The partition is not lumpable (Definition~\ref{def:lumpable}), since a time $t$ after the outcome $N$ the outcome $S$ occurs at the rate
$\eta|\braket{S|e^{-iHt}|N}|^2=\eta\sin^2\!\alpha\,\sin^2(\omega t/2)$, which oscillates with $t$. The record is the MPS of Eq.~\eqref{eq:II-mps} with bond dimension $2$, the spin, with the matrices
$A^\varnothing=\one_{\mathrm{sys}}-i\big(H-\frac i2\eta\big)\dd t$, $A^N=\sqrt{\eta\,\dd t}\,\ket N\bra N$ and $A^S=\sqrt{\eta\,\dd t}\,\ket S\bra S$ and
the boundary vector $\ket N$. The bias is the MPS of Eq.~\eqref{eq:III-Wmps} with bond dimension $2$, the register, with the matrices
$W^\varnothing=\mathrm{diag}\big(e^{-\frac s2\cos\alpha\,\dd t},e^{\frac s2\cos\alpha\,\dd t}\big)$, $W^N=\ket0\big(\bra0+\bra1\big)$ and
$W^S=\ket1\big(\bra0+\bra1\big)$, which write the latest outcome into the register, and the boundary vector $\ket0$. Their product of Sec.~\ref{sec:product} has the matrices $A_W^\varnothing=A^\varnothing\otimes W^\varnothing$
and $A_W^{(\mu,m)}=A^\mu\otimes W^\mu\ket m\bra m$, for $\mu=N,S$ and $m=0,1$, with bond dimension $4$, and the leading left eigenvector
$G=G_0\otimes\ket0\bra0+G_1\otimes\ket1\bra1$ of its transfer map, $\sum_xA_W^{x\dagger}GA_W^x=e^{\theta\,\dd t}G$, has the blocks
\begin{equation}\label{eq:spin-G}
  G_m=\int_0^\infty\!\dd t\,e^{-r_mt}\,e^{iHt}Je^{-iHt}
\end{equation}
with $J\coloneqq\eta\big(g(\ket N)\ket N\bra N+g(\ket S)\ket S\bra S\big)$,
$r_0=\eta+\theta+s\cos\alpha$ and $r_1=\eta+\theta-s\cos\alpha$, which are both at least
$\eta\cos^2\!\alpha>0$, since $\theta\ge0$ and, by trajectories that stay in $\Reach_0$ or in $\Reach_1$, $\theta\ge|s|\cos\alpha-\eta\sin^2\!\alpha$. The optimal instrument is the left canonical
form $A_L^x=e^{-\theta\,\dd t/2}G^{1/2}A_W^xG^{-1/2}$ of Eq.~\eqref{eq:III-gauge}.
The two numbers $g(\ket N)=\braket{N|G_0|N}$ and $g(\ket S)=\braket{S|G_1|S}$ then fix $\theta$ as the largest root of
$(\theta+s\cos\alpha+\eta p_0)(\theta-s\cos\alpha+\eta p_1)=\eta^2p_0p_1$, with $p_0\coloneqq\frac12\sin^2\!\alpha\,\omega^2/(r_0^2+\omega^2)$
and $p_1\coloneqq\frac12\sin^2\!\alpha\,\omega^2/(r_1^2+\omega^2)$, which is the leading eigenvalue of $\LbiasW$.

For $\omega=\eta$, $\alpha=\pi/3$ and $s=-\eta$, the post-measurement state lies in $\Reach_0$ for a fraction
$\frac12\big(1-\theta'(s)/\cos\alpha\big)\approx0.95$ of the time instead of the unbiased case $\frac12$, by Eq.~\eqref{eq:II-mean}, see Figs.~\ref{fig:spin}(b) and~\ref{fig:spin}(c), whereas post-selection reaches these trajectories only with a number of repeated experiments
that grows as $e^{TI(-\theta'(s))}\approx e^{0.15\eta T}$, by Eq.~\eqref{eq:II-ldp}.

By Eqs.~\eqref{eq:law} and~\eqref{eq:spin-G}, the
waiting time until the next outcome decays as $e^{-r_0t}$ after an outcome $N$ and as $e^{-r_1t}$ after an outcome $S$, up to oscillations at the
precession frequency, with $r_0\approx0.79\,\eta$ and $r_1\approx1.79\,\eta$ for $s=-\eta$, instead of $\eta$ for both without the bias. The optimal
rates favour the outcome $N$, which returns the state to $\Reach_0$, as $g$ is larger on $\Reach_0$ than on $\Reach_1$, and as $g$ varies along each
$\Reach_m$, they depend on the precession phase and not only on $m$.
\section{Example: Free fermions with gain and loss}\label{sec:ff}

\begin{figure}[t]
\centering
\includegraphics{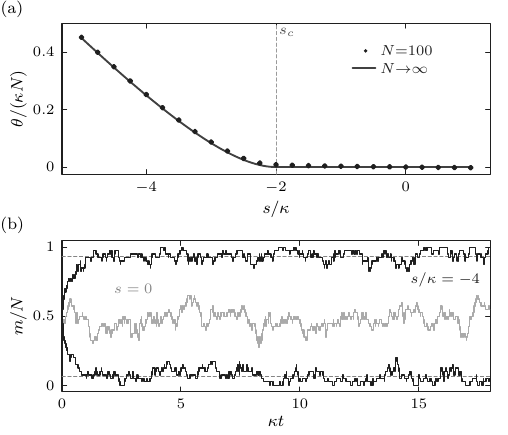}
\caption{Free fermions with gain and loss at rate $\kappa$ on a ring of $N$ sites, biased by Eq.~\eqref{eq:IV-bias}. (a)~$\theta/(\kappa N)$ at $N=100$ (dots) and its large-$N$ limit~\eqref{eq:IV-mf} (line), with the transition at $s_c=-2\kappa$. (b)~The particle density $m/N$ held by the register along trajectories of the optimal instrument at $N=40$, started at half filling, one at $s=0$ (grey) and two at $s=-4\kappa$ (black), which settle near $\hat\nu$ (dashed).}
\label{fig:ff}
\end{figure}

\subsection{The model and its post-measurement states}\label{sec:ff-model}

Consider spinless fermions on a ring of $N$ sites with the hopping Hamiltonian
\begin{equation*}
  H\coloneqq\sum_{a,b=1}^{N}h_{ab}\,c_a^\dagger c_b,\qquad h_{j,j+1}=h_{j+1,j}=-J,
\end{equation*}
where all other $h_{ab}$ vanish and $c_{N+1}\equiv c_1$. Every site loses and gains particles at the same rate $\kappa$, i.e., the jump operators
are $L_{j-}\coloneqq\sqrt\kappa\,c_j$ and $L_{j+}\coloneqq\sqrt\kappa\,c_j^\dagger$, with $j=1,\dots,N$. The outcome $j-$ is a loss and the
outcome $j+$ a gain at site $j$. Since $c_j^\dagger c_j+c_jc_j^\dagger=\one_{\mathrm{sys}}$,
\begin{equation*}
  \Gamma=\kappa N\,\one_{\mathrm{sys}},\qquad \Heff=H-\tfrac i2\kappa N\,\one_{\mathrm{sys}} ,
\end{equation*}
so that the evolution for outcome $\varnothing$ is unitary, $\ket{\psi(t)}=e^{-iHt}\ket\psi$, and the outcomes other than $\varnothing$ occur at the
total rate $\kappa N$ in every state: with $n_j\coloneqq c_j^\dagger c_j$ and $\langle n_j\rangle_\psi\coloneqq\braket{\psi|n_j|\psi}$, a loss at $j$
occurs at the rate $r_{j-}=\kappa\langle n_j\rangle_\psi$ and a gain at the rate $r_{j+}=\kappa(1-\langle n_j\rangle_\psi)$.

We start from a state $\ket{\psi_0}$ with a definite particle number $m_0$, e.g., the vacuum. As $H$ commutes with $\hat N\coloneqq\sum_jn_j$, while
$c_j$ and $c_j^\dagger$ lower and raise $\hat N$ by one, every reachable state of Definition~\ref{def:reach} has a definite particle number
$m(t)$, which every gain raises and every loss lowers by one.

\subsection{Biasing by the particle number}\label{sec:ff-spin}

We bias by the particle number of the post-measurement state, with
\begin{equation}\label{eq:IV-bias}
  f(\ket\psi)\coloneqq f(m),\qquad f(m)\coloneqq\frac{(m-N/2)^2}{N},
\end{equation}
for $\ket\psi\in\Reach$ with $\hat N\ket\psi=m\ket\psi$, so that $W=e^{-sA_T}$ with $A_T=\int_0^Tf\big(m(t)\big)\,\dd t$. For $s<0$ the bias favours
trajectories along which the particle number strays far from half filling, whereas for $s>0$ it favours those along which it stays close to it.
As $f$ is nonlinear,
$A_T$ is not the time integral of a linear combination of the numbers of gains and losses.

The sets $\Reach_m\coloneqq\{\ket\psi\in\Reach:\hat N\ket\psi=m\ket\psi\}$, $m\in\Mem\coloneqq\{0,\dots,N\}$, form a partition for the bias
(Definition~\ref{def:partition}) with $\varphi_{j-}(m)=m-1$ and $\varphi_{j+}(m)=m+1$, since $f$ is constant on each $\Reach_m$ and $e^{-iHt}$
conserves $\hat N$; the values $\varphi_{j-}(0)$ and $\varphi_{j+}(N)$ do not matter, since $c_j$ and $c_j^\dagger$ annihilate the states with $0$
and $N$ particles. Hence Assumption~\ref{ass:finite} holds, with $\pi(\ket\psi)$ the particle number of $\ket\psi$. As the subspaces of
Eq.~\eqref{eq:Sm} we take the mutually orthogonal sectors $\mathcal S_m$ of $m$ particles, on which $\sum_jL_{j-}^\dagger L_{j-}=\kappa\hat N$ and
$\sum_jL_{j+}^\dagger L_{j+}=\kappa(N-\hat N)$ equal $\kappa m$ and $\kappa(N-m)$, so that the partition is lumpable (Definition~\ref{def:lumpable}),
with $q(m\to m-1)=\kappa m$ and $q(m\to m+1)=\kappa(N-m)$. By Theorem~\ref{thm:lumpable}, $m$ performs the irreducible random walk of the
Ehrenfest model~\cite{Ehr07}, as if each particle were lost, and each empty site filled, at rate $\kappa$, and $\theta$ and
$g$ follow from Eq.~\eqref{eq:label}, which reads
\begin{align}
  \theta\,g(m)={}&\kappa m\big(g(m-1)-g(m)\big)-s\,f(m)\,g(m)\label{eq:IV-label}\\
  &+\kappa(N-m)\big(g(m+1)-g(m)\big),\nonumber
\end{align}
for $m=0,\dots,N$, where $g(-1)=g(N+1)\coloneqq0$.

As for any master equation of independent two-state units~\cite{ADHR94}, conjugation by the
diagonal matrix with entries $\binom Nm^{-1/2}$ makes $\Lambda_s$ symmetric, with the off-diagonal entries $\kappa\sqrt{m(N-m+1)}$ of
$2\kappa S^x$ for a spin $N/2$ and the diagonal $-\kappa N-s\,f(m)$, where $S^z=m-N/2$ on the basis state $\ket m$ of the register. As
$f(m)=(S^z)^2/N$, $-\theta(s)$ is the ground-state energy of the Lipkin--Meshkov--Glick (LMG)
Hamiltonian~\cite{LMG65}
\begin{equation}\label{eq:IV-LMG}
  H_{\rm LMG}\coloneqq-2\kappa S^x+\frac sN\big(S^z\big)^2+\kappa N .
\end{equation}

\subsection{The transition at $s_c=-2\kappa$}\label{sec:ff-transition}

At large $N$, the ground-state energy of the LMG model is given by spin coherent states~\cite{BJ83,DV05}. In the coherent state in the $xz$
plane with $S^z\approx N(\nu-\frac12)$, i.e., with the fraction $\nu$ of occupied sites, $S^x\approx N\sqrt{\nu(1-\nu)}$, so that
\begin{align}
  \lim_{N\to\infty}\frac{\theta(s)}N&=\max_{\nu\in[0,1]}\Big(2\kappa\sqrt{\nu(1-\nu)}-\kappa-s\big(\nu-\tfrac12\big)^2\Big)\nonumber\\
  &=\begin{cases}0, & s\ge-2\kappa,\\[2pt] -\dfrac{(s+2\kappa)^2}{4s}, & s<-2\kappa,\end{cases}\label{eq:IV-mf}
\end{align}
where the maximum is attained at $\hat\nu=\frac12$ for $s\ge-2\kappa$ and at $\hat\nu=\frac12\pm\big(\frac14-\kappa^2/s^2\big)^{1/2}$ for $s<-2\kappa$, see Fig.~\ref{fig:ff}. For
$s\ge-2\kappa$ the transverse field dominates, the spin points along $x$, and the biased trajectories stay at half filling, with $|m-\frac N2|$
of order $\sqrt N$ as in typical unbiased trajectories. For $s<-2\kappa$ the coupling $\frac sN(S^z)^2$ dominates, the spin turns towards $\pm z$,
and the particle--hole symmetry $m\to N-m$ of the bias is broken: the biased trajectories carry an extensive excess of particles or of holes,
$|m-\frac N2|\approx N|\hat\nu-\frac12|$, see Fig.~\ref{fig:ff}(b). This is the ground-state transition of the LMG model, with its mean-field
exponents~\cite{BJ83,DV04}: just below $s_c=-2\kappa$, $\theta/N\approx(s-s_c)^2/(8\kappa)$ and $|\hat\nu-\frac12|\approx\big((s_c-s)/(4\kappa)\big)^{1/2}$.

The transition has a simple meaning. In Eq.~\eqref{eq:IV-mf}, $\kappa\big(\sqrt\nu-\sqrt{1-\nu}\big)^2=\kappa-2\kappa\sqrt{\nu(1-\nu)}$ is the cost,
per site and unit time, of holding the density at $\nu$ against the independent gains and losses, and $-s(\nu-\frac12)^2$ is the reward of the
bias. Close to half filling, with $\nu=\frac12+\delta$, the cost is $\approx2\kappa\delta^2$ and, for $s<0$, the reward is $|s|\delta^2$, so that the
biased trajectories leave half filling once the reward wins, at $s_c=-2\kappa$.

\subsection{The optimal instrument}\label{sec:ff-instrument}

Since $g(\ket\psi)=g(m)$, the optimal rates of Eq.~\eqref{eq:optrates} keep the evolution $e^{-iHt}$ for outcome $\varnothing$ and rescale the rates
of loss and gain, which become
\begin{equation*}
\begin{split}
  r^g_{j-}(\ket\psi)&=\kappa\,\langle n_j\rangle_\psi\,\frac{g(m-1)}{g(m)},\\
  r^g_{j+}(\ket\psi)&=\kappa\,\big(1-\langle n_j\rangle_\psi\big)\,\frac{g(m+1)}{g(m)},
\end{split}
\end{equation*}
for $\ket\psi\in\Reach_m$. They depend on $\ket\psi$ through $\langle n_j\rangle_\psi$, which the chain supplies through the Born rule, and through $m$,
which is fixed by the record.

By Theorem~\ref{thm:lumpable}(iii), with $G_m=g(m)P_m$, the instrument of Theorem~\ref{thm:instrument} acts on the space $\mathcal V$ of chain $\otimes$ register,
with the register basis $\ket0,\dots,\ket N$, and has the jump operators
\begin{align}
  L^G_{j-,m}&=\Big(\kappa\,\frac{g(m-1)}{g(m)}\Big)^{1/2}c_j\otimes\ket{m-1}\bra m,\label{eq:IV-instrument}\\
  L^G_{j+,m}&=\Big(\kappa\,\frac{g(m+1)}{g(m)}\Big)^{1/2}c_j^\dagger\otimes\ket{m+1}\bra m,\nonumber
\end{align}
for $1\le m\le N$ and $0\le m\le N-1$, respectively, and the effective Hamiltonian
$H^G_{\rm eff}=\sum_m\big(H-\frac i2(\kappa N+\theta+s\,f(m))\big)\otimes\ket m\bra m$. Its register holds the particle number of the chain.

Identifying $\mathcal V$ with the Fock space by $\ket\phi\otimes\ket m\mapsto\ket\phi$, for $\ket\phi\in\mathcal S_m$, the $L^W_{j\pm,m}$ become
$\sqrt\kappa\,c_jP_m$ and $\sqrt\kappa\,c_j^\dagger P_m$, whose sums over $m$, $\sqrt\kappa\,c_j$ and $\sqrt\kappa\,c_j^\dagger$, generate all operators on the
Fock space, so that $\LbiasW$ generates irreducible CP maps, as Theorem~\ref{thm:absorb} requires.
As the sectors $\mathcal S_m$ are mutually orthogonal and span the Fock space, Theorem~\ref{thm:absorb} absorbs the register into the chain. As
$c_j$ and $c_j^\dagger$ map each sector into a single sector, and the reachable states have a definite particle number, the jump operators of
Eq.~\eqref{eq:absorb} for different $m$ can be added into one for each outcome $j\pm$, so that the optimal dynamics is generated by the Lindbladian
on the chain alone with the hopping Hamiltonian $H$ and the jump operators
\begin{equation*}
\begin{split}
  L^g_{j-}&\coloneqq\sqrt\kappa\,c_j\Big(\frac{g(\hat N-1)}{g(\hat N)}\Big)^{1/2},\\
  L^g_{j+}&\coloneqq\sqrt\kappa\,c_j^\dagger\Big(\frac{g(\hat N+1)}{g(\hat N)}\Big)^{1/2}.
\end{split}
\end{equation*}
These jump operators act as follows. For $s<0$, $g(m)$ grows away from half filling, so that above half filling the gains are enhanced and
the losses suppressed, and conversely below. For $s>s_c$, this push is weaker than the drift $\kappa(N-2m)$ of the unbiased gains and losses
towards half filling, and $m$ stays near $\frac N2$. For $s<s_c$, it wins: half filling becomes unstable, and $m$ settles near one of the two
densities $\hat\nu$ of Eq.~\eqref{eq:IV-mf}, see Fig.~\ref{fig:ff}(b).
\section{Conclusion and outlook}\label{sec:concl}

In this work, we have established a quantum instrument approach to realise the rare trajectories of monitored quantum dynamics without post-selection. Using tensor networks, we represent the measurement record as an MPS whose auxiliary space is the system, the bias as another MPS over the record whose auxiliary space is the register, and the optimal instrument is the left canonical form of the corresponding MPS. 

We have illustrated these results with two examples. For a single spin that precesses in a magnetic field and is measured at random times along an axis at an angle to the field, biased by its spin component along the field. And for free fermions with gain and loss, biased by a function of the particle number. In addition, we have argued the entanglement transition of the Majorana chain of Appendix~\ref{sec:chain} under a technical assumption. 

Several problems remain open. Many biases of interest have no partition into finitely many sets $\Reach_m$, e.g., the entanglement entropy of a generic post-measurement state. Such biases require an infinite register, and the natural approximation is to truncate its bond dimension, as in MPS approaches to dynamical large deviations of classical models~\cite{BG19,CBG22}, with the variational principle of Sec.~\ref{sec:optimal} providing a measure of what such a truncation loses. Another direction is to design an experiment on current platforms with mid-circuit measurements, in which the optimal instrument, an adaptive circuit whose register is set by classical feedback, is compared with the exponential cost of post-selecting trajectories that stay mixed.

\begin{acknowledgments}
We have used Claude and ChatGPT to assist in developing and preparing the
manuscript. All results, proofs, calculations, and references were independently verified by the authors, who take full responsibility for the content of the manuscript.
\end{acknowledgments}

\appendix
\section{Majorana chain}\label{sec:chain}

\subsection{The model and its post-measurement states}\label{sec:pairings}

Following Refs.~\cite{NS20,LB20}, we consider $2N$ Majorana operators $\gamma_1,\dots,\gamma_{2N}$, i.e., $N$ fermionic modes, and measure
each of the $2N-1$ parities of neighbouring Majoranas,
\begin{equation}\label{eq:V-P}
  P_j\coloneqq i\gamma_j\gamma_{j+1},\qquad j=1,\dots,2N-1,
\end{equation}
at the times of an independent Poisson process of rate $1$. As a monitored dynamics, it has $H=0$ and the jump operators
$L_{j,\sigma}=\Pi^\sigma_j\coloneqq\frac12(\one_{\mathrm{sys}}+\sigma P_j)$, $\sigma=\pm1$, i.e., the outcome $(j,\sigma)$ is the outcome $\sigma$ of a
measurement of $P_j$. Since $\sum_\sigma\Pi^\sigma_j=\one_{\mathrm{sys}}$, $\Gamma=(2N-1)\one_{\mathrm{sys}}$ and $\Heff$ is
a multiple of the identity, so that the evolution for outcome $\varnothing$ does not change the post-measurement state.

\begin{figure}[t]
\centering
\includegraphics{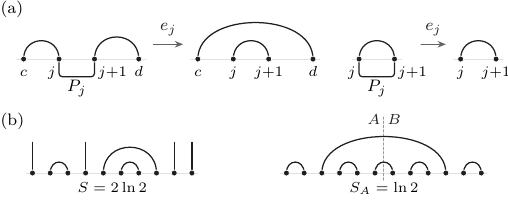}
\caption{Pairings of the Majoranas. (a)~A measurement of the parity $P_j=i\gamma_j\gamma_{j+1}$ (bracket) acts on the pairing (arcs) by the move $e_j$ of Eq.~\eqref{eq:V-move}, which rewires two arcs (left) or, if $(j,j+1)$ is an arc, leaves the pairing unchanged (right). (b)~A mixed state with $2k$ unpaired Majoranas (vertical lines) has $S=k\ln2$ (left), and a pure state has $S_A=\frac{\ln2}2n_{AB}$, with $n_{AB}$ the number of arcs across the cut (dashed line, right).}
\label{fig:chain}
\end{figure}

The post-measurement states are described by pairings of the Majoranas~\cite{NS20}. A pairing $\alpha$ is a non-crossing perfect matching of
$\{1,\dots,2N\}$, i.e., a set of $N$ arcs $(a,b)$ with $a<b$, no two of which cross ($a<c<b<d$). There are $C_N=(2N)!/\big(N!(N+1)!\big)$ such
pairings, and we denote their set by $\mathrm{NC}_{2N}$. For a pairing $\alpha$ and signs $\sigma_{ab}=\pm1$, the $N$ commuting operators
$\sigma_{ab}\,i\gamma_a\gamma_b$, $(a,b)\in\alpha$, stabilise a unique state $\ket{\psi(\alpha,\sigma)}$.

Every reachable state of Definition~\ref{def:reach}, started in the state with $P_1=P_3=\dots=P_{2N-1}=1$, i.e., with the pairing $\alpha_0\coloneqq\{(1,2),(3,4),\dots,(2N-1,2N)\}$ and all signs $+1$, is of
this form~\cite{NS20}. If $(j,j+1)\in\alpha$, the
outcome of $P_j$ is deterministic and the state is unchanged. Otherwise, $P_j$ anticommutes with the two stabilisers that contain $\gamma_j$ and
$\gamma_{j+1}$, both outcomes have probability $\frac12$, and the measurement replaces these two stabilisers by $\sigma\,i\gamma_j\gamma_{j+1}$ and
the product of both with $\sigma\,i\gamma_j\gamma_{j+1}$, which is $\pm i\gamma_c\gamma_d$ with $c$ and $d$ the former partners of $j$ and $j+1$. The pairing becomes
\begin{equation}\label{eq:V-move}
  e_j\cdot\alpha\coloneqq\alpha\setminus\{(j,c),(j+1,d)\}\cup\{(j,j+1),(c,d)\}
\end{equation}
(each pair denoting an unordered arc), and we set $e_j\cdot\alpha\coloneqq\alpha$ when $(j,j+1)\in\alpha$. Starting from
$\alpha_0$, the pairing thus performs a random walk on $\mathrm{NC}_{2N}$, the raise-and-peel
model~\cite{dGNPR04}, which applies each move $e_j$ at rate $1$, whatever the signs and the outcomes, see Fig.~\ref{fig:chain}(a).

\subsection{Biasing by the pairing: the XXZ chain}\label{sec:xxz}

We bias by a function of the pairing, $f\big(\ket{\psi(\alpha,\sigma)}\big)=f(\alpha)$. The sets $\Reach_\alpha$ of the reachable states with the
pairing $\alpha$, $\alpha\in\mathrm{NC}_{2N}$, form a partition for the bias (Definition~\ref{def:partition}) with the maps
$\varphi_{(j,\sigma)}(\alpha)=e_j\cdot\alpha$, since $f$ is constant on each $\Reach_\alpha$, the evolution for outcome $\varnothing$ does not change the
state, and both outcomes of the measurement of $P_j$ lead to the pairing $e_j\cdot\alpha$. Hence Assumption~\ref{ass:finite} holds, with $\Mem$ the set
$\mathrm{NC}_{2N}$ of $C_N$ pairings, so that $m$ is the pairing $\alpha$, which the register holds. As subspaces of Eq.~\eqref{eq:Sm} we take every $\mathcal S_\alpha$ to be the
sector of total fermion parity $P_1P_3\cdots P_{2N-1}=1$, which contains the initial state and is mapped into itself by every $\Pi^\sigma_j$, since the
total fermion parity, proportional to $\gamma_1\gamma_2\cdots\gamma_{2N}$, commutes with every $P_j$. Since $\sum_\sigma\Pi^\sigma_j=\one_{\mathrm{sys}}$, the
measurement of $P_j$ occurs at the total rate $1$ from every state, so that the partition is lumpable (Definition~\ref{def:lumpable}), with
$q(\alpha\to\alpha')$ the number of $j$ with $e_j\cdot\alpha=\alpha'$. By Theorem~\ref{thm:lumpable}, $\theta$ and $g$ follow from the $C_N\times C_N$
matrix
\begin{equation}\label{eq:V-Lambda}
  \Lambda_s=\Lambda-s\,\hat f,\qquad \Lambda=\sum_{j=1}^{2N-1}\big(E_j-\one\big),
\end{equation}
where $E_j\coloneqq\sum_\alpha\ket{e_j\cdot\alpha}\bra\alpha$ on the register. Since every pairing is reached from every other by moves $e_j$
(Appendix~\ref{sec:C-moves}), the walk is irreducible.

The rate matrix $\Lambda$ is minus the Temperley--Lieb Hamiltonian at loop weight $1$ on link patterns~\cite{NS20,dGNPR04}. For the mixed states of Sec.~\ref{sec:purity}, the register holds a pattern with $2k$ defects, i.e., unpaired points under no arc, $\Lambda$ is
block triangular in $k$, and its diagonal block $\Lambda^{(k)}$ acts on a standard module of the Temperley--Lieb algebra~\cite{TL71,Mar91,RSA14},
see Eq.~\eqref{eq:C-en}. Following Morin-Duchesne~\cite{MD11}, we map the
patterns to spin states by
\begin{equation}\label{eq:V-D}
  D(\alpha)\coloneqq U\bigotimes_{(a,b)\in\alpha}(\mathfrak q^{\frac12}\ket{\uparrow_a\downarrow_b}+\mathfrak q^{-\frac12}\ket{\downarrow_a\uparrow_b})\bigotimes_d\ket{\uparrow_d},
\end{equation}
where the second product is over the defects $d$, with $\mathfrak q\coloneqq e^{i\pi/3}$, so that $\mathfrak q+\mathfrak q^{-1}=1$, and $U\coloneqq\prod_{j\ \rm odd}\sigma^z_j$, i.e., the map of Ref.~\cite{MD11} at $w=\mathfrak q^{1/2}$,
multiplied by $U$. This map is injective, takes the patterns with $2k$ defects into the sector $S^z_{\rm tot}=k$, and satisfies
$H_{\rm XXZ}D=-D\Lambda^{(k)}$ there~\cite{MD11}, where
\begin{align}
  H_{\rm XXZ}\coloneqq{}&2\sum_{j=1}^{2N-1}\big(S^x_jS^x_{j+1}+S^y_jS^y_{j+1}+\tfrac12S^z_jS^z_{j+1}\big)\label{eq:V-XXZ}\\
  &+\tfrac34(2N-1)-\tfrac{i\sqrt3}{2}\big(S^z_1-S^z_{2N}\big)\nonumber
\end{align}
is the open XXZ chain at anisotropy $\Delta=\frac12$ with the boundary terms of Refs.~\cite{ABBBQ87,PS90}.

\subsection{Biasing by the von Neumann entropy}\label{sec:purity}

We start the chain in the maximally mixed state $\one_{\mathrm{sys}}/2^N$, which measurements purify~\cite{GH20,NS20}, and take every
$\mathcal S_\alpha=\mathbb C^{2^N}$. For a mixed post-measurement state $\rho$, Definitions~\ref{def:partition} and~\ref{def:lumpable} and
Theorem~\ref{thm:lumpable} apply with $\braket{\psi|\cdot|\psi}$ replaced by $\Tr(\cdot\,\rho)$, as one sees by purifying $\rho$ with a
reference system that is not measured, and, as $\rho$ is stabilised by the operators $\sigma_{ab}\,i\gamma_a\gamma_b$,
\begin{equation}\label{eq:V-mixed}
  \rho=2^{-N}\prod_{(a,b)\in\alpha}\big(\one_{\mathrm{sys}}+\sigma_{ab}\,i\gamma_a\gamma_b\big)
\end{equation}
at all times, where $\alpha$ is a pattern with $2k$ defects, i.e., unpaired Majoranas, so that the von Neumann entropy of $\rho$ is $k\ln2$, see Fig.~\ref{fig:chain}(b). A
measurement of $P_j$ acts on $\alpha$ by $e_j$, except that two adjacent defects are paired, i.e., purification is the pairwise annihilation of
defects.

We bias by the von Neumann entropy, $A_T\coloneqq\int_0^TS\big(\rho(t)\big)\,\dd t$ with $S(\rho)\coloneqq-\Tr(\rho\ln\rho)$, i.e., $f(\alpha)=k\ln2$,
and the patterns form a lumpable partition, as the pairings do. As $k$ never increases, $\theta$ is the largest over $k$ of the leading eigenvalues
of the blocks of $\Lambda_s$ with $2k$ defects, see Eq.~\eqref{eq:V-thetaN}, attained at $k=\hat k(s)$, the number of pairs of defects that the biased trajectories keep, and,
by the map $D$, each of them is minus an eigenvalue of $H_{\rm XXZ}-2hS^z_{\rm tot}$ in the sector $S^z_{\rm tot}=k$, with
$h\coloneqq-\frac12s\ln2$, i.e., the entropy bias is a magnetic field on the XXZ chain.

\begin{figure}[t]
\centering
\includegraphics{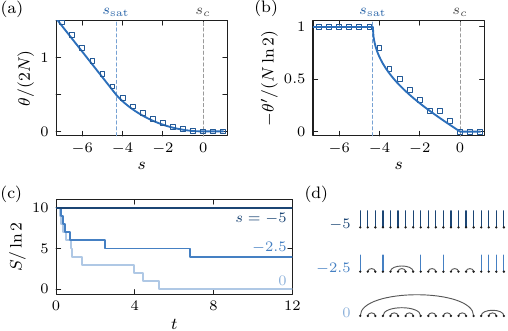}
\caption{Biasing by the von Neumann entropy $S$, starting from $\one_{\mathrm{sys}}/2^N$, with $s_c=0$ and $s_{\rm sat}=-3/\ln2$ (dashed). (a)~$\theta/(2N)$ at $2N=20$ (squares) and its large-$N$ limit~\eqref{eq:V-limit} (line). (b)~The mean entropy $-\theta'$ divided by $N\ln2$, i.e., $\hat k/N$ at $2N=20$ (squares), and its limit $2m_z(h)$ (line). (c)~$S/\ln2$ along trajectories of the optimal instrument at $2N=20$ for $s=0,-2.5,-5$. (d)~Their patterns at $t=12$, with the unpaired Majoranas as vertical lines.}
\label{fig:entropy}
\end{figure}

Let $E_0(L;k)$ be the ground-state energy of the open chain $H_{\rm XXZ}$ of Eq.~\eqref{eq:V-XXZ} on $L=2N$ sites in the sector
$S^z_{\rm tot}=k$, i.e., its eigenvalue with the smallest real part, which is real, as we checked for all $k$ and $L\le14$. By the map $D$, $\theta\le-\min_k\big(E_0(L;k)-2hk\big)$ for every $L$. We note that $\mathfrak q=e^{i\pi/3}$ is a root of unity, a special
point at which, e.g., $H_{\rm XXZ}$ is not diagonalisable, so that results for generic $\mathfrak q$ need not carry over. To
evaluate $\theta$ at large $N$, we therefore make one assumption, namely that for $s\le0$, i.e., $h\ge0$, $\theta$ is given by this ground-state energy,
and that the open chain has the ground-state energy of the periodic chain up to a term of order one,
\begin{align}
  \theta(s)&=-\min_{0\le k\le N}\big(E_0(L;k)-2hk\big)\label{eq:V-open}\\
  &=-L\big(2\varepsilon_h+\tfrac34\big)+O(1),\nonumber
\end{align}
where $\varepsilon_h$ is the ground-state energy per site of the periodic chain
$\sum_{j=1}^L(S^x_jS^x_{j+1}+S^y_jS^y_{j+1}+\frac12S^z_jS^z_{j+1}-hS^z_j)$, which is solved by the Bethe ansatz~\cite{YY66,Tak99,Fra17}.

\begin{figure}[t]
\centering
\includegraphics{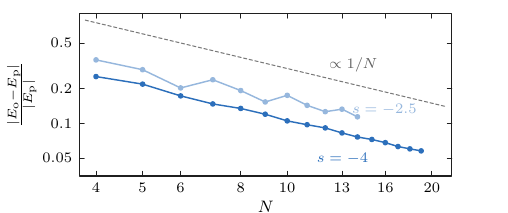}
\caption{Relative difference of the ground-state energies $E_{\rm o}$ of the open chain of Eq.~\eqref{eq:V-XXZ}, with its complex boundary terms, and $E_{\rm p}$ of the periodic chain, in the field $h=-\frac12s\ln2$ at $s=-2.5$ and $s=-4$, by exact diagonalisation of both, against the number $N$ of sites, with a line $\propto1/N$ (dashed).}
\label{fig:surface}
\end{figure}

Equation~\eqref{eq:V-open} is exact in the two limits of the field: at $s=0$, $\theta=0$ (Appendix~\ref{sec:C-moves}) and $\varepsilon_0=-\frac38$,
with no term $O(1)$, and in the saturated phase, where the maximum in Eq.~\eqref{eq:V-thetaN} is attained at $k=N$, all moves annihilate the empty
pattern, so that $\theta=-(L-1)+hL$, while $\varepsilon_h=\frac18-\frac h2$, the term $O(1)$ being the energy $1$ of the bond missing from the open chain. Fig.~\ref{fig:surface} shows the the relative
difference of the ground-state energies of the open and the periodic chain that falls as $1/N$.

By the Bethe ansatz of the periodic chain~\cite{YY66,Tak99,Fra17},
\begin{align}
  &\varepsilon_h=\tfrac18-\tfrac h2+\int_{-B}^{B}a_1(\lambda)\,\eta(\lambda)\,\dd\lambda,\label{eq:V-BA}\\
  &\eta(\lambda)+\int_{-B}^{B}a_2(\lambda-\lambda')\,\eta(\lambda')\,\dd\lambda'=h-\tfrac{\sqrt3\pi}2\,a_1(\lambda),\nonumber
\end{align}
where $a_n(\lambda)\coloneqq\sin\frac{n\pi}3/\big(\pi(\cosh2\lambda-\cos\frac{n\pi}3)\big)$ and the dressed energy $\eta$ vanishes at $\lambda=\pm
B$~\cite{YY66,Tak99}. At $h=0$, $B=\infty$ and $\varepsilon_0=-\frac38$~\cite{YY66,RS01}, and $B$ decreases with $h$ to $0$ at $h=1+\Delta$, where
the chain is fully polarised and $s=s_{\rm sat}\coloneqq-3/\ln2$. Hence, assuming Eq.~\eqref{eq:V-open}, see Fig.~\ref{fig:entropy}(a),
\begin{equation}\label{eq:V-limit}
  \lim_{N\to\infty}\frac{\theta(s)}{2N}=\begin{cases}0,&s\ge0,\\-2\big(\varepsilon_h-\varepsilon_0\big),&s_{\rm sat}<s<0,\\-\frac12s\ln2-1,&s\le s_{\rm sat}.\end{cases}
\end{equation}
Between $s_{\rm sat}$ and $0$, Eq.~\eqref{eq:V-BA} has no closed-form solution. Near the two ends, the magnetisation per site,
$m_z(h)=-\partial_h\varepsilon_h$, is $m_z(h)\simeq\chi h$, with the zero-field susceptibility $\chi=1/(\sqrt3\pi)$, and
$m_z(h)\simeq\frac12-\frac1\pi\sqrt{2(1+\Delta-h)}$, as the flipped spins are dilute and free fermions~\cite{YY66,Tak99,Fra17}.
By Eq.~\eqref{eq:II-mean}, the von Neumann entropy of the biased trajectories is $-\theta'(s)=\hat k\ln2$, and $\hat k/N\to2m_z(h)$, the
magnetisation curve of the chain, see Fig.~\ref{fig:entropy}(b). The biased ensemble therefore has a purification transition at $s_c=0$. For $s\ge0$, the biased trajectories purify, as the typical unbiased ones do, whereas for every $s<0$ they keep, at large $N$, an entropy proportional to $N$. 

Biasing pure post-measurement states, started with the pairing $\alpha_0$ and $N$ even, by their half-chain entanglement entropy instead gives a
volume law. This entropy is $S_A=\frac{\ln2}2n_{AB}(\alpha)$, with $n_{AB}(\alpha)$ the number of arcs of $\alpha$ across the middle cut~\cite{NS20}, see
Fig.~\ref{fig:chain}(b). Without the measurement of $P_N$, the parity across the cut, the reduced states of the two halves are of the
form of Eq.~\eqref{eq:V-mixed}, each with entropy $S_A$. Without the term $E_N-\one$, $\Lambda_s$ is therefore block triangular in $n_{AB}$, its diagonal blocks are those of two
such halves, each biased by $\frac s2$ times its entropy, and its leading eigenvalue is $2\theta_N(\frac s2)$, with $\theta_N$
the value of $\theta$ above for $N$ Majoranas. As $E_N$ has non-negative entries, restoring the term $E_N-\one$ lowers the leading eigenvalue by at most
$1$, so that $\theta(s)\ge2\theta_N(\frac s2)-1\ge-1$, where $-1$ is the cost of never measuring $P_N$. As $\theta$ is convex with $\theta(0)=0$, the mean
of $S_A$ in the biased ensemble, $-\theta'(s)$, is then at most $1/s$ for $s>0$ and at least $\theta(s)/|s|$ for $s<0$. Hence the half-chain
entanglement entropy of the biased trajectories changes from an area law for $s>0$ through the logarithmic growth of the typical trajectories at
$s=0$~\cite{NS20} to a volume law for $s<0$, where, assuming Eq.~\eqref{eq:V-open}, $\theta$ grows linearly with $N$.
\section{Temperley--Lieb}\label{app:tl}\label{sec:C-moves}

This appendix gives the statements on link patterns used for the Majorana chain of Appendix~\ref{sec:chain}. To our
knowledge, the reachability of patterns with defects has not appeared in the literature in this form.

For even $L$ and $0\le k\le L/2$, let $\mathrm{NC}^{(k)}_L$ be the set of non-crossing matchings of $L-2k$ of the points $1,\dots,L$ whose $2k$ unmatched
points, called defects, lie under no arc. We have $\mathrm{NC}^{(0)}_L=\mathrm{NC}_L$.
The segments between consecutive defects, as well as those before the first and after the last defect, are matched within themselves and hence
have even length. On $\mathbb C^{\mathrm{NC}^{(k)}_L}$, let
\begin{equation}\label{eq:C-en}
  E_j\alpha\coloneqq\begin{cases}\alpha,&(j,j+1)\in\alpha,\\0,&j,j+1\ \text{both defects},\\e_j\cdot\alpha,&\text{otherwise},\end{cases}
\end{equation}
where $e_j\cdot\alpha$ pairs $j$ with $j+1$ and joins their former partners as in Eq.~\eqref{eq:V-move} or, if one of $j,j+1$ is a defect and
the other is paired with $c$, turns $c$ into a defect. This is the standard module of the Temperley--Lieb algebra at loop weight $1$ with $2k$
through-lines~\cite{TL71,Mar91,RSA14}, and we write $H^{\rm TL}\coloneqq\sum_{j=1}^{L-1}(\one-E_j)$. The three cases are the moves of a mixed
post-measurement state of the form of Eq.~\eqref{eq:V-mixed} with a deterministic outcome, with the annihilation of two unpaired Majoranas (set to zero), and
with an outcome of probability $\frac12$, so that $-H^{\rm TL}$ on $\mathbb C^{\mathrm{NC}^{(k)}_L}$ is the block $\Lambda^{(k)}$ of the rate matrix
$\Lambda$, which loses probability through the annihilations. As $-H^{\rm TL}$ has non-negative off-diagonal entries, its leading eigenvalue
$\lambda_L(k)$, i.e., the eigenvalue with the largest real part, is real~\cite{BP94}.

We first show that the moves connect the patterns. Two statements are needed, namely that any two patterns of $\mathrm{NC}^{(k)}_L$ are
connected by a sequence of moves $e_j$, none of which is applied to two adjacent defects, and that every pattern of every
$\mathrm{NC}^{(k)}_L$ is reached from the empty pattern ($k=L/2$) by moves $e_j$.

Consider first the full patterns, $k=0$. Starting from any $\alpha$, we apply $e_1,e_3,\dots,e_{L-1}$ in turn. Once the arcs
$(1,2),\dots,(2l-1,2l)$ are present, the partners of $2l+1$ and $2l+2$ lie to their right, so that $e_{2l+1}$ creates $(2l+1,2l+2)$ without
affecting the earlier arcs, and in this way $\alpha_0$ is reached. Conversely, let $|\alpha|\coloneqq\sum_{(a,b)\in\alpha}(b-a)$ be the total length of
the arcs, which is minimal only at $\alpha_0$. If $\alpha\ne\alpha_0$, we take an arc $(a,b)$ of minimal length $b-a\ge3$, so that the arcs
inside it have length $1$ and $(a+1,a+2)\in\alpha$. Replacing $(a,b),(a+1,a+2)$ by $(a,a+1),(a+2,b)$ gives a pattern $\alpha'$ with
$|\alpha'|=|\alpha|-2$ and $e_{a+1}\cdot\alpha'=\alpha$, and hence, by induction on $|\alpha|$, $\alpha$ is reached from $\alpha_0$.

With defects present, the moves also transport them. If $d$ is a defect and $d+1$ is paired with $c$, then $c>d+1$, since otherwise $d$
would lie under $(c,d+1)$, and $e_d$ moves the defect from $d$ to $c$. Starting from the rightmost defect, every defect is moved to the right
in this way until the defects occupy $L-2k+1,\dots,L$, after which the full pattern on $1,\dots,L-2k$ is brought to the dimers
$(1,2),(3,4),\dots$ as for $k=0$. Similarly, if $d$ is a defect and $(d-2,d-1)$ is an arc, then $e_{d-1}$ moves the defect from $d$ to $d-2$.
Given a target $\alpha$ with defects $d_1<\dots<d_{2k}$, we move the first defect to the left over dimers to $d_1$, then the second to $d_2$,
and so on, the parities being matched because the segments have even length. This gives dimers in every segment, and the case $k=0$ within each
segment then gives $\alpha$. Since no step applies $e_j$ to two adjacent defects, this proves the first statement. For the second, we start from
the empty pattern, pair adjacent defects so as to fill every segment of $\alpha$ with dimers, and use the case $k=0$ within each segment.

Moreover, $\lambda_L(k)\le0$, as the columns of $-H^{\rm TL}$ on $\mathbb C^{\mathrm{NC}^{(k)}_L}$ sum to minus the
number of $j$ for which $j,j+1$ are both defects, which is non-positive~\cite{BP94}.

For the bias by the von Neumann entropy, ordered by $k$, $\Lambda_s$ is block triangular with the diagonal blocks
$-H^{\rm TL}-sk\ln2$ on $\mathbb C^{\mathrm{NC}^{(k)}_L}$, so that its eigenvalues are those of the blocks, and as every pattern is reached from the
empty one, in which the chain starts, every block contributes to $Z_T$. Hence, for every $L$,
\begin{equation}\label{eq:V-thetaN}
  \theta(s)=\max_{0\le k\le L/2}\big(\lambda_L(k)-s\,k\ln2\big),
\end{equation}
attained at $k=\hat k(s)$. Since the columns of $\Lambda$ on $\mathrm{NC}_L$ sum to zero, $\lambda_L(0)=0$, so that $\theta(s)=0$ for $s\ge0$.

\renewcommand{\href}[2]{#2}
\bibliography{mps_ld,newrefs}

\end{document}